\documentclass[11pt]{scrartcl} 
\ifx\pdfminorversion\undefined
\else
\fi

\usepackage{cite}
\usepackage{amsmath,amssymb,amsfonts}
\usepackage{algorithmic}
\usepackage{graphicx}
\usepackage{textcomp}
\usepackage{lineno,hyperref,float,amsfonts,epsfig,graphics,booktabs,multirow,cite,stfloats}
\usepackage{bm}
\usepackage{comment}
\usepackage{amsthm}
\newcommand{\tabincell}[2]{\begin{tabular}{@{}#1@{}}#2\end{tabular}}

\theoremstyle{definition}
\newtheorem{myDef}{Definition}
\theoremstyle{plain}
\newtheorem{myTheo}{Theorem}
\newtheorem{myCoro}{Corollary}
\newtheorem{myLemma}{Lemma}
\theoremstyle{remark}
\newtheorem{myRemark}{Remark}
\newtheorem{myProp}{Proposition}

\usepackage[caption=false]{subfig}

\title{ Adaptive robust tracking control with active learning for linear systems with ellipsoidal bounded uncertainties }

\author{Xuehui~Ma\footnote{School of Automation and Information Engineering, Xi'an University of Technology, Xi'an, China. E-mail: xuehui.yx@gmail.com, qianfc@xaut.edu.cn} \and Shiliang~Zhang\footnote{Department of Informatics, University of Oslo, Oslo, Norway. E-mail: shilianz@ifi.uio.no, yushuaili@ieee.org} \and Yushuai Li$^\dagger$ \and Fucai Qian$^\ast$ \and Tingwen Huang\footnote{T \& M University at Qatar, Doha 23874, Qatar. E-mail: tingwen.huang@qatar.tamu.edu\\This work was supported by the National Natural Science Foundation of China (Grant no. 62073259).}  }

\begin{document}

\maketitle

\begin{abstract}
This paper is concerned with the robust tracking control of linear uncertain systems, whose unknown system parameters and disturbances are bounded within ellipsoidal sets. We propose an adaptive robust control that can actively learn the ellipsoid sets. Particularly, the proposed approach utilizes the recursive set-membership state estimation in learning the ellipsoidal sets, aiming at mitigating uncertainties in the system control. Upon the learned sets representing the recognized uncertainties, we construct a robust control with one-step prediction for system output tracking. In deriving an optimized control law, we reformulate the optimization objective into a second-order cone programming problem that can be solved in a computationally friendly way. To further stimulate the active learning of uncertainties over the control procedures, we enrich the information used for the learning by maximizing the volume of the ellipsoid set, supposed to lead to increased learning accuracy and accelerated uncertainty reduction. To verify our approach, we conduct numerical simulations to compare the fixed-ellipsoidal-set robust control with ours, and investigate the positive effect of the designed active learning in the uncertain system control process.
\end{abstract}

\textit{keywords:} robust control, adaptive control, bounded uncertainties, ellipsoidal set, set-membership filter.

\section{Introduction}
The control of practical systems inevitably involves handling uncertainties, since the system dynamic in its nature can be unclear, intervened by disturbances in the surroundings, or varying in an unknown manner within the control horizon~\cite{petersen2014robust,wang2017adaptive,lewis2017optimal}. This renders the issue of the control of uncertain systems, where the challenge lies in uncertainty identification and reduction while maintaining the control process over time~\cite{8666991,mesbah2018stochastic}. In this work, we look into solutions for the control of uncertain systems with unknown but bounded system parameters and disturbances.

Stochastic adaptive control is an established methodology for the control of uncertain systems~\cite{aastrom2012introduction,filatov2000survey,lewis2017optimal}. Works on this method assume uncertainties in the system follow known probability distributions, such as Gaussian and asymmetric Laplace distribution~\cite{mesbah2018stochastic,li2008optimal,ma2022adaptive}, and adaptively identify the system dynamics during the control process. However, the statistical properties of uncertainties are quite often unknown or irregular in practice. Instead, only bounds of the possible magnitude of the uncertainties are available~\cite{becis2008state,filatov2003design,9780027}, making stochastic adaptive control inefficient in such cases. 

Differing from stochastic adaptive control, robust control seeks a control law that guarantees the worst-case performance within the uncertainty boundaries~\cite{9969863,1231253,lee1997worst,9865123}. Robust control strategies assume a fixed bounded scale for uncertainties and derive the robust control law by minimizing the maximum performance cost within the fixed boundaries. Nevertheless, the probability of uncertain systems falling into the worst-case situation is extremely low in practical scenarios. As a result, the worst-case control strategy is over-conservative and sacrifices the tracking performance in guaranteeing the robustness of the control. 

Investigations have tried to combine the advanced features of stochastic adaptive control and robust control for uncertain systems. Such efforts aim to identify the uncertainty boundaries by utilizing the information from system observations, where the range of the uncertainty boundaries is reduced and updated over time. Several works~\cite{9304303,9780027,10155796,SASFI2023111169,HEYDARI2021109672,LU2023110959,10179989,Wabersich5688,9440698,Jianglin5235,9029693,9143777,YANG2023111056,Lingyi6237} have integrated parameter recognition or state estimation into robust controller design for systems with bounded uncertainties. Amongst those works, the set-membership identification stands out as an efficient approach, which can recognize parameters for uncertainties represented by bounded sets and reduce the level of conservation as is in robust control with fixed uncertainty boundaries. Lorenzen et al. utilized the set-membership identification to learn the unknown parameters in the development of robust regulation of linear systems ~\cite{lorenzen2019robust,LORENZEN20173313}. Köhler et al. integrated the set-membership filter based parameter estimation into the design of tube-based model predictive control for nonlinear uncertain systems~\cite{kohler2021robust}. Recently, Parsi et al. proposed an MPC control for uncertain systems with dual properties where the unknown system parameters are actively learned by set-membership identification~\cite{9780027,9304303}. 

A major issue in set-membership estimation is its subjection to the use of polytopic sets, which are used to describe uncertainties. The complexity of the algorithm associated with polytopic sets grows dramatically with system dimensions, jeopardizing its availability in a broader range of applications. Numbers of studies have turned to ellipsoids instead of polytopes~\cite{lorenzen2019robust} when it comes to set-membership estimation. Set-membership estimation based on ellipsoidal outer bounding was first introduced by Schweppe and Witsenhausen~\cite{1098790,1098995}, and has been well developed over the last decades~\cite{maksarov1996state,becis2008state,WANG2019337,9437722}. Set-membership estimation with ellipsoids provides computational friendliness in identifying unknown parameters during system control. Furthermore, the use of ellipsoids enables (i) a Kalman-filter liked mechanism that can be recursively conducted and integrated into the control process, and (ii) the uncertainty that can be more easily quantified by the trace of shape matrix compared with polytopic approaches~\cite{becis2008state,lorenzen2019robust,parsi2022scalable}.

Ellipsoidal set based set-membership estimation has been utilized to handle bounded uncertainties in several works on optimal control. Kothare et al.~\cite{kothare1996robust} considered both the parameter and structure uncertainties in system control, and solved the min-max control law using linear matrix inequality, yet they did not update the uncertainty set during the control. Qian et al.~\cite{qian2010adaptive} proposed an optimal tracking control for uncertain systems with parameters and disturbances bounded in ellipsoid sets, where the ellipsoid sets are updated along the control process. However, their work only considered the uncertainty of control matrix for the system, while they ignored that of the state parameter matrix. Parsi et al.~\cite{parsi2022scalable} and Schwenkel et al.~\cite{9765740} proposed an ellipsoidal tube based model predictive control with ellipsoidal sets learned iteratively. Nevertheless, their methods merely take into account the control performance and neglect the quality of the learned uncertainty sets. 

In this work, we emphasize the importance of effective recognition and reduction of uncertainties in the control of uncertain systems, represented by the quality of the learned ellipsoidal sets during the set-membership estimation. An efficient way to enhance the learning of ellipsoidal sets is to enrich the information used for this learning. Particularly, the maximization of the volume of the ellipsoidal set at each learning iteration can stimulate the system to acquire more information to feed the learning. Yet the volume maximization leads to the conflict between the control and the learning, a.k.a. the dual control problem~\cite{filatov2000survey}. I.e., the maximum ellipsoidal volume implies the burden to maintain the highest level of uncertainty during the control. Substantial numbers of studies are dedicated to sub-optimal methods in solving this dilemma in the field of dual control~\cite{filatov2000survey,mesbah2018stochastic}, while the developed solutions have not been well transferred to solve robust control issues. Here we leverage the bi-criterion dual control approach~\cite{filatov2004adaptive,MA2022157,9189668} to address the conflict between learning and control, and contribute to our design of a novel adaptive robust control strategy that can actively learn uncertainties during the control.

This paper integrates the ellipsoidal set based set-membership estimation into the optimal robust tracking control for linear uncertain systems, which are corrupted by bounded parameters and additive disturbances. Through this effort, we endow the control system with active learning toward uncertainties in an adaptive way. The main contributions of this paper are:

\begin{itemize}
	\item {we propose an adaptive robust tracking control formulation for systems corrupted by unknown but bounded parameters and exogenous disturbances, where we use ellipsoidal sets to describe parameter uncertainties and to bound process disturbances.}
	\item {we reform uncertain systems into the formulation with unknown parameters and disturbances. We utilize the ellipsoidal set estimation to learn the uncertainties during the control, and derive an optimal robust control law. To the best of our knowledge, we are the first to integrate the ellipsoidal set learning into robust control and derive an optimal robust control with one-step prediction.}
	\item {To stimulate the system's active learning of uncertainties, we tailor the one-step prediction control by adding another optimization objective, i.e., maximizing the volume of the updated ellipsoidal set. We leverage the bi-criterion approach to coordinate the conflict between the tracking performance and the learning quality, and derive the adaptive robust control law with full-fledged learning capacity.}
\end{itemize}
 
This paper is organized as follows. We list notations and definitions for our work following this introduction. Section~\ref{section2} formulates the tracking control problem with uncertainties bounded by ellipsoidal sets. The ellipsoidal set based parameter learning is shown in~\ref{section3}, and the designed optimal robust control with one-step prediction is presented in section~\ref{section4}. Section \ref{section5} details the derivation of the control law with active learning. In section \ref{section6}, numerical simulations, comparisons, and analyses are conducted to demonstrate the proposed approach. We conclude this work in Section~\ref{section7}.

\quad

\noindent \textit{Notations and definitions:}
\begin{myDef}
	$\bm{P}\succ 0 $ means the matrix $\bm{P} $ is positive semi-definite, and $\bm{P} \succeq 0 $ means $\bm{P} $ is positive definite. 
\end{myDef}

\begin{myDef}
	$\mathcal{E}(\bm{P},\bm{a}) \triangleq \{ \bm{x} \in \mathbb{R}^n | (\bm{x}-\bm{a})^T \bm{P}^{-1} (\bm{x}-\bm{a}) \leq 1 \} $ is an ellipsoid in $\mathbb{R}^n$, where $\bm{a} \in \mathbb{R}^n $ is the center of the ellipsoid, matrix $\bm{P}\in \mathbb{R}^{n \times n} $ determines the shape, size and orientation of the ellipsoid subject to $\bm{P}\succ 0 $. 
\end{myDef}


\begin{myDef}
	Let $\bm{x_a} \in \mathcal{E}(\bm{P_a},\bm{a}) $ and $\bm{x_b} \in \mathcal{E}(\bm{P_b},\bm{b})$, where $\bm{a}$ and $\bm{b}$ $ \in \mathbb{R}^n$, $\bm{P_a}$ and $\bm{P_b} \in \mathbb{R}^{n \times n}$, $\bm{P_a}$ and $\bm{P_b}\succ 0$. Then the sum of the two ellipsoidal set is $\mathcal{E}(\bm{P_a},\bm{a}) \oplus \mathcal{E}(\bm{P_b},\bm{b}) \triangleq \{ \bm{x} \in \mathbb{R}^n | \bm{x}=\bm{x_a}+\bm{x_b}\}$, and the intersection of the two ellipsoidal set is $\mathcal{E}(\bm{P_a},\bm{a}) \cap \mathcal{E}(\bm{P_b},\bm{b}) \triangleq \{ \bm{x} \in \mathbb{R}^n | \bm{x}=\bm{x_a} \cap \bm{x_b}\}$.
\end{myDef}

\begin{myProp}\label{prop1}
	\cite{maksarov1996state} $\mathcal{E}(\bm{P_a},\bm{a}) \oplus \mathcal{E}(\bm{P_b},\bm{b}) \subseteq \mathcal{E}(\bm{P_c},\bm{c})$, where $\bm{c}=\bm{a}+\bm{b}$, $\bm{P_c}=(\tau^{-1}+1)\bm{P_a}+(\tau+1)\bm{P_b}$, and the scalar parameter $0<\tau<\infty$. The optimal value of $\tau$ minimizing the size of $\mathcal{E}(\bm{P_c},\bm{c})$ satisfies 
	\begin{equation}\label{para1}
		\begin{aligned}
			\sum_{i=1}^{n} \frac{1}{\lambda_i(\bm{P_a}\bm{P_b}^{-1})+\tau}=\frac{n}{\tau(\tau+1)},
		\end{aligned}
	\end{equation}
where $\lambda_i(\bm{P_a}\bm{P_b}^{-1})$ denotes the eigenvalue $i$ of the corresponding matrix $\bm{P_a}\bm{P_b}^{-1}$. 
\end{myProp}

\begin{myProp}\label{prop2}
	\cite{maksarov1996state} $\mathcal{E}(\bm{P_a},\bm{a}) \cap \mathcal{E}(\bm{P_b},\bm{b}) \subseteq \mathcal{E}(\bm{P_c},\bm{c})$, where $\bm{c}=\bm{a}+\bm{L}(\bm{b}-\bm{a})$, $\bm{L}=\bm{P_a}(\bm{P_a}+\rho^{-1}\bm{P_b})^{-1}$, $\bm{P_c}=\beta(\rho)((\bm{I}-\bm{L})\bm{P_a}(\bm{I}-\bm{L})^T+\rho^{-1}\bm{L}\bm{P_b}\bm{L}^T)$, $\beta(\rho)=1+\rho-(\bm{b}-\bm{a})^T(\bm{P_a}+\rho^{-1}\bm{P_b})^{-1}(\bm{b}-\bm{a})$, and $\beta(\rho)>0$ for all scalar parameter $\rho \geq 0$. The optimal value of $\rho$ minimizing the size of $\mathcal{E}(\bm{P_c},\bm{c})$ satisfies 
	\begin{equation}\label{para2}
		\begin{aligned}
			\sum_{i=1}^{n} \frac{\lambda_i (\bm{P_a}\bm{P_b}^{-1})}{1+\rho \lambda_i (\bm{P_a}\bm{P_b}^{-1})}=n\frac{\beta'(\rho)}{\beta(\rho)},
		\end{aligned}
	\end{equation}
	where $\beta'(\rho)=1-(\bm{b}-\bm{a})^T(\bm{P_a}+\rho^{-1}\bm{P_b})^{-1}\rho^{-2}\bm{P_b}(\bm{P_a}+\rho^{-1}\bm{P_b})^{-1}(\bm{b}-\bm{a})$, and $\lambda_i(\bm{P_a}\bm{P_b}^{-1})$ denotes the eigenvalue $i$ of the corresponding matrix $\bm{P_a}\bm{P_b}^{-1}$. 
\end{myProp}

\section{Problem Statement }\label{section2}
Consider an uncertain discrete-time linear system 
\begin{equation}\label{sys1}
	\begin{aligned}
			\bm{x}(k+1) = \bm{A}(\bm{\alpha})\bm{x}(k)+\bm{B}(\bm{\beta})\bm{u}(k)+\bm{\omega}(k),\\
			k=1,2, \cdots, N,
	\end{aligned}
\end{equation}
where $k$ is the discrete time index, $\bm{x}(k) \in \mathbb{R}^n $ is the system state and assumed to be observable, $\bm{u}(k) \in \mathbb{R}^m$ denotes the control input, and $\bm{\omega}(k) \in \mathbb{R}^l $ is the disturbance that bounded in ellipsoids and can be represented as  
\begin{equation}\label{disturbellip}
	\begin{aligned}
		\bm{\omega}(k) \in \mathcal{E}(\bm{R}(k),\bm{0}),
	\end{aligned}
\end{equation}
where $\bm{R}(k) \succ 0$. The system parameter $\bm{A}\in \mathbb{R}^{n \times n} $ and $\bm{B}\in \mathbb{R}^{n \times m} $ are supposed to be bounded but unknown, and could be described by unknown and unmeasurable parameter vectors $\bm{\alpha}\in \mathbb{R}^{r} $ and $\bm{\beta}\in \mathbb{R}^{s} $, represented as $\bm{A}(\bm{\alpha})$ and $\bm{B}(\bm{\beta})$, respectively. For ease of exposition, we define the uncertain matrix $\bm{A}(\bm{\alpha})$ and $\bm{B}(\bm{\beta})$ as affine functions of the parameter vectors $\bm{\alpha}=[\alpha_1, \alpha_2, \cdots, \alpha_r]^T$ and $\bm{\beta}=[\beta_1, \beta_2, \cdots, \beta_s]^T$, respectively, which can be written as
\begin{equation}\label{paraA}
	\begin{aligned}
		\bm{A}(\bm{\alpha})=\bm{A}_0+\sum_{i=1}^{r}\bm{A}_i \alpha_i
	\end{aligned}
\end{equation}
\begin{equation}\label{paraB}
	\begin{aligned}
		\bm{B}(\bm{\beta})=\bm{B}_0+\sum_{i=1}^{s}\bm{B}_i \beta_i ,
	\end{aligned}
\end{equation}
where $\bm{A}_i$ and $\bm{B}_i$ are given nominal matrices. To simplify the expression, we define the parameter vector $\bm{\theta}$ to aggregate $\bm{\alpha}$ and $\bm{\beta}$ as
\begin{equation}\label{paratheta}
	\begin{aligned}
		\bm{\theta}& = [\theta_1, \theta_2, \cdots, \theta_{r+s}]^T \\
		&= [\alpha_1, \alpha_2, \cdots, \alpha_r, \beta_1, \beta_2, \cdots, \beta_s]^T .
	\end{aligned}
\end{equation}

\begin{figure}[htbp]
	\centering
	\includegraphics[width=0.5\textwidth]{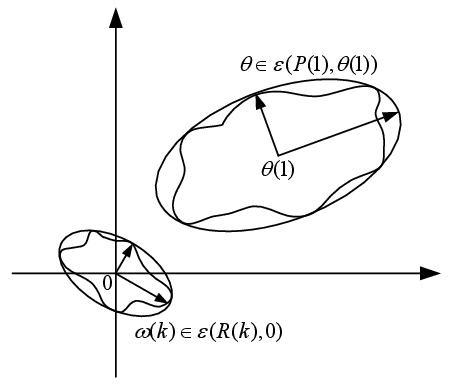}\\
	\caption{The geometric interpretation of ellipsoidal sets.}\label{ellipsoidset}
\end{figure}

We assume that the initial value of $\bm{\theta} $ is bounded in the following ellipsoid
\begin{equation}
	\begin{aligned}
		\bm{\theta} \in \mathcal{E}(\bm{P}(1),\bm{\theta}(1)),
	\end{aligned}
\end{equation}
where $\bm{P}(1)\succ 0 $ and $\bm{\theta}(1)$ are known.  Fig.~\ref{ellipsoidset} shows the geometric interpretation of the ellipsoidal sets for uncertain parameters $\bm{\theta}$ and disturbance $\bm{\omega}$ with a two-dimensional example.
The output to be tracked is given by 
\begin{equation}\label{sys2}
	\begin{aligned}
		\bm{y}(k) = \bm{C}(k)\bm{x}(k),
	\end{aligned}
\end{equation}
where $\bm{y}(k) \in \mathbb{R}^l $, and the output matrix $\bm{C}(k) \in \mathbb{R}^{l \times n} $ are supposed to be known,

The objective of this paper is to design a robust controller that minimizes the deviation of the system output from a reference sequence under worst-case scenarios. We mathematically formulate this objective as $(\mathcal{P})$:
\begin{equation}\label{the control objective}
	\begin{aligned}
		(\mathcal{P}): \quad &  \min_{\bm{u}(k)} \max_{ \{\bm{\theta}(k), \bm{\omega}(k)\} } \sum_{k=1}^{N} \| \bm{y}(k+1)-\bm{y}_r(k+1) \|_1 \\
		 \quad s.t. \quad & \bm{x}(k+1) = \bm{A}(\bm{\alpha})\bm{x}(k)+\bm{B}(\bm{\beta})\bm{u}(k)+\bm{\omega}(k),\\
		 \quad \quad \quad \quad & \bm{y}(k) = \bm{C}(k)\bm{x}(k),\\
		 \quad \quad \quad \quad & \bm{u}_{min} \leq \bm{u}(k) \leq \bm{u}_{max},\\
		 \quad \quad \quad \quad & k=1, 2, \cdots, N ,
	\end{aligned}
\end{equation}
where $\bm{y}_r(k)$ is the reference sequence for the system output, $\bm{u}_{min}$ and $\bm{u}_{max}\in \mathbb{R}^m$ are the lower and upper bounds of the control signal, respectively, and the performance index for the control is 
\begin{equation}\label{optperformanceindex}
	\begin{aligned}
		J = \max_{ \{\bm{\theta}(k), \bm{\omega}(k)\} } \sum_{k=1}^{N} \| \bm{y}(k+1)-\bm{y}_r(k+1) \|_1 \\
	\end{aligned}
\end{equation}

According to the principle of optimality~\cite{milito1982innovations}, the robust optimal control sequence for the objective $(\mathcal{P})$ could be obtained by dynamic programming algorithm shown below:
\begin{equation}\label{J_N}
	\begin{aligned}
		J_N^*[\bm{\mathfrak{I}}(N)] = 0,
	\end{aligned}
\end{equation}
\begin{equation}\label{bellmaneq}
	\begin{aligned}
		J_k[\bm{u}(k), \bm{\mathfrak{I}}(k)] = \max_{ \{\bm{\theta}(k), \bm{\omega}(k)\} } \left\lbrace D_{k+1}[\bm{\mathfrak{I}}(k+1)] + J_{k+1}^*[\bm{\mathfrak{I}}(k+1)]\right\rbrace ,
	\end{aligned}
\end{equation}
\begin{equation}\label{bellmaneq_derive1}
	\begin{aligned}
		D_{k+1}[\bm{\mathfrak{I}}(k+1)] = \|\bm{y}(k+1)-\bm{y}_r(k+1)\|_1,
	\end{aligned}
\end{equation}
\begin{equation}\label{bellmaneq_derive2}
	\begin{aligned}
		J_k^*[\bm{\mathfrak{I}}(k)] = \min_{\bm{u}(k)} J_k[\bm{u}(k), \bm{\mathfrak{I}}(k)],
	\end{aligned}
\end{equation}
\begin{equation}\label{bellmaneq_derive3}
	\begin{aligned}
		\bm{u}^*(k) = \arg\min_{\bm{u}(k)} J_k[\bm{u}(k), \bm{\mathfrak{I}}(k)],
	\end{aligned}
\end{equation}
where $\bm{\mathfrak{I}}(k)$ is the information state that contains the initial information and past inputs \& states as $\bm{\mathfrak{I}}(k)=\{\bm{u}(1), \cdots, \bm{u}(k-1), \bm{x}(1), \cdots, \bm{x}(k)\}$. 

\begin{myRemark} \label{remark1}
	From~(\ref{J_N}) to (\ref{bellmaneq_derive3}), we observe that the optimal robust control law $\bm{u}^*$ achieves a compromise between the immediate control objective and the future control objective, represented by the one-step cost function $D_{k+1}[\bm{\mathfrak{I}}(k+1)]$ and the cost function $J_{k+1}^*[\bm{\mathfrak{I}}(k+1)]$, respectively. Solving such an optimization in $(\mathcal{P})$ by directly applying dynamic programming is difficult due to the prohibitive cost of computation and ram storage. An alternative solution is to omit $J_{k+1}^*[\bm{\mathfrak{I}}(k+1)]$ in (\ref{bellmaneq}) and derive an immediate optimal robust control law $\bm{u}_I(k)$~\cite{milito1982innovations}. Such a control law can be derived by minimizing the one-step prediction error $D_{k+1}[\bm{\mathfrak{I}}(k+1)]$. Nevertheless, this solution only cares about improving the immediate control performance, yet prevents the attainment of overall control objective of minimizing $J$ in (\ref{optperformanceindex}), leading to a myopic control law that is far from the optimal robust control law~\cite{1137553}. We are interested in this work to reformulate the one-step prediction cost function, aiming to actively acquire more information needed to drive the immediate control closer to an optimal control solution with computational feasibility.
\end{myRemark}

\begin{myRemark}\label{remark2}
	Assume the uncertainty set $\bm{\theta}$ is fixed at any iteration $k$ during the control, then it is easy to derive an optimal robust controller by solving the problem $(\mathcal{P})$ in~(\ref{the control objective}). We can achieve this solution by minimizing the worst performance under the fixed uncertainty set. However, the worst-performance case hardly happens in practical systems, and the corresponding solution is more than often over-conservative with a large fixed uncertainty set and results in low tracking performance~\cite{9780027,lorenzen2019robust}. Such a disadvantage motivates us to design a control strategy that can learn the uncertainty set rather than using a fixed one. We anticipate the prospective solution to reduce the uncertainty of the parameter $\bm{\theta}$ in deriving the robust control law and improving the system tracking performance.
\end{myRemark}

\section{Learning of the Uncertainty}\label{section3}

In this section, we detail our design of uncertainty learning in the control of uncertain systems, and explain how our design can reduce the uncertainties and contribute to the control.

Uncertainties induced by outer environment or inner system widely exist in practical systems, and can be classified into irreducible and reducible uncertainty~\cite{li2008optimal}. In the problem ($\mathcal{P}$) in (\ref{the control objective}), we assume that the bounded disturbance $\bm{\omega}(k)$ is of a random nature and is regarded as an irreducible uncertainty, and the unknown constant parameter $\bm{\theta}$ be a reducible uncertainty and can be learned from information state $\bm{\mathfrak{I}}(k+1)$. 

This section reforms the system (\ref{sys1}) and (\ref{sys2}) into the observation equation with parameter $\bm{\theta}$, and utilizes the observation to learn the uncertainty. The learning process updates the ellipsoid from a prior $\mathcal{E}(\bm{P}(k),\hat{\bm{\theta}}(k))$ to a posterior $\mathcal{E}(\bm{P}(k+1),\hat{\bm{\theta}}(k+1))$ using the observation $\bm{x}(k)$, as detailed in the following.

Based on the system described in (\ref{sys1}) and (\ref{sys2}), we can define an auxiliary function $\bm{H}(k+1)\in \mathbb{R}^l $ as
\begin{equation}\label{auxi1}
	\begin{aligned}
		\bm{H}(k+1) = &\bm{y}(k+1)-\bm{C}(k+1)\bm{A}_0(k)\bm{x}(k)\\
		&-\bm{C}(k+1)\bm{B_0}\bm{u}(k),
	\end{aligned}
\end{equation}
and define the vector $\bm{\phi}(k) \in \mathbb{R}^{l \times s} $ as 
\begin{equation}\label{auxi2}
	\begin{aligned}
		\bm{\phi}(k) = &[\bm{C}(k+1)\bm{A}_1(k)\bm{x}(k), \bm{C}(k+1)\bm{A}_2(k)\bm{x}(k), \\ 
		&\cdots, \bm{C}(k+1)\bm{A}_r(k)\bm{x}(k), \bm{C}(k+1)\bm{B_1}\bm{u}(k), \\ 
		& \bm{C}(k+1)\bm{B_2}\bm{u}(k), \cdots,	\bm{C}(k+1)\bm{B_s}\bm{u}(k)], \\
	\end{aligned}
\end{equation}
and disturbance $\bm{\upsilon}(k) \in \mathbb{R}^l$ as
\begin{equation}\label{auxi3}
	\begin{aligned}
		\bm{\upsilon}(k) = \bm{C}(k+1)\bm{\omega}(k).
	\end{aligned}
\end{equation}

According to the definition in $\ref{paratheta}$, $(\ref{auxi1})$, $(\ref{auxi2})$, $(\ref{auxi3})$, and the system dynamics $(\ref{sys1})$, $(\ref{sys2})$, the defined auxiliary function $\bm{H}(k+1)$ can be rewritten as
\begin{equation} \label{Hfunction}
	\begin{aligned}
		\bm{H}(k+1) & = \bm{\phi}(k) \bm{\theta}+\bm{\upsilon}(k),
	\end{aligned}
\end{equation}
where $\bm{H}(k+1)$ and $\bm{\phi}(k)$ can be calculated by the observed system output $\bm{y}(k+1)$, $\bm{y}(k)$ and the control signal $\bm{u}(k)$. 
The ellipsoidal set of $\bm{\upsilon}(k)$ satisfies 
\begin{equation}
	\begin{aligned}
		\bm{\upsilon}(k) \in \mathcal{E}(\bm{Q}(k),\bm{0}) ,
	\end{aligned}
\end{equation}
where the matrix $\bm{Q}(k)$ can be calculated by  $\bm{Q}(k)=\bm{C}(k+1)\bm{R}(k)\bm{C}^T(k+1)$.

The obtained function $\bm{H}(k+1)$ confines $\bm{\theta}$ to the observation ellipsoid 
\begin{equation}
	\begin{aligned}
		\mathcal{E}(k+1) = \{ \bm{\theta} \in \mathbb{R}^s | (\bm{H}(k+1)-\bm{\phi}(k)\bm{\theta})^T \\
		\bm{Q}^{-1}(k) (\bm{H}(k+1)-\bm{\phi}(k)\bm{\theta}) \leq 1 \},
	\end{aligned}
\end{equation}
and this ellipsoidal set can be formatted as
\begin{equation}
	\begin{aligned}
		& \mathcal{E}(\bm{\phi}^{-1}(k)\bm{Q}(k)\bm{\phi}^{-T}(k), \bm{\phi}^{-1}(k)\bm{H}(k+1) )\\
		& = \{ \bm{\theta} \in \mathbb{R}^s | (\bm{\theta}-\bm{\phi}^{-1}(k)\bm{H}(k+1))^T \bm{\phi}^T(k) \bm{Q}^{-1}(k) \\
		& \quad \quad \quad \quad \quad  \bm{\phi}(k)(\bm{\theta}-\bm{\phi}^{-1}(k)\bm{H}(k+1)) \leq 1 \}.
	\end{aligned}
\end{equation}


Initially, the prior ellipsoid at the iteration $k+1$ is equal to the posterior ellipsoid at $k$, demonstrated as $\mathcal{E}(\bm{P}(k+1|k),\hat{\bm{\theta}}(k+1|k))=\mathcal{E}(\bm{P}(k),\hat{\bm{\theta}}(k))$. After applying the control signal $\bm{u}(k)$ to the system, we can observe the output $\bm{y}(k+1)$ and obtain the auxiliary function $\bm{H}(k+1)$. Then we use $\bm{y}(k+1)$ and $\bm{H}(k+1)$ to update the prior ellipsoid $\mathcal{E}(\bm{P}(k+1|k),\hat{\bm{\theta}}(k+1|k))$, so as to learn the posterior ellipsoid $\mathcal{E}(\bm{P}(k+1|k+1),\hat{\bm{\theta}}(k+1|k+1))$. Below is how the posterior ellipsoid is learned.

With the observations at the $(k+1)$-th iteration, $\bm{\theta}$ is confined in the intersection between the prior ellipsoid $\mathcal{E}(\bm{P}(k+1|k),\hat{\bm{\theta}}(k+1|k))$ and the observation ellipsoid $\mathcal{E}(\bm{\phi}^{-1}(k)\bm{Q}(k)\bm{\phi}^{-T}(k), \bm{\phi}^{-1}(k)\bm{H}(k+1) )$
\begin{equation}
	\begin{aligned}
		\bm{\theta} \in \mathcal{E}(\bm{P}(k+1|k),\hat{\bm{\theta}}(k+1|k)) \cap  \\
		 \mathcal{E}(\bm{\phi}^{-1}(k)\bm{Q}(k)\bm{\phi}^{-T}(k),\bm{\phi}^{-1}(k)\bm{H}(k+1) ) .\\	
	\end{aligned}
\end{equation}

Since the intersection is not in the form of an ellipsoid, we use the proposition \ref{prop2} to derive an ellipsoid that encloses the intersection, and regard the derived ellipsoid as the posterior ellipsoid
\begin{equation}\label{posterior}
	\begin{aligned}
		\mathcal{E}(\bm{P}(k+1|k+1),\hat{\bm{\theta}}(k+1|k+1)) \supseteq \mathcal{E}(\bm{P}(k+1|k), \\ \hat{\bm{\theta}}(k+1|k)) \cap \mathcal{E}(\bm{\phi}^{-1}(k)\bm{Q}(k)\bm{\phi}^{-T}(k),  \bm{\phi}^{-1}(k)\bm{H}(k+1) ).\\	
	\end{aligned}
\end{equation}

\begin{figure}[htbp]
	\centering
	\includegraphics[width=0.6\textwidth]{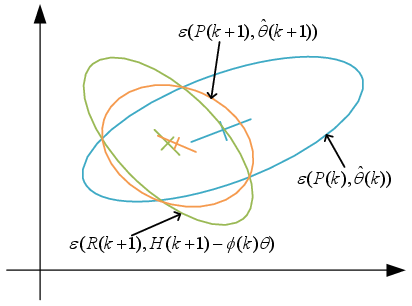}\\
	\caption{The geometric interpretation of posterior ellipsoid set.}\label{posteriorellipsoid}
\end{figure}

Fig.~\ref{posteriorellipsoid} shows the geometric interpretation of the posterior ellipsoid set. The solution to obtain the posterior ellipsoid set is detailed in Theorem~\ref{theo1} in the following.

\begin{myTheo}\label{theo1}
	\cite{maksarov1996state} $\forall~0\leq\rho<\infty$, the posterior ellipsoid centre $\hat{\bm{\theta}}(k+1|k+1)$ and the shape matrix $\bm{P}(k+1|k+1)$ satisfying (\ref{posterior}) are shown as
	\begin{equation}
		\begin{aligned}
			\hat{\bm{\theta}}(k+1|k) = \hat{\bm{\theta}}(k|k)
		\end{aligned}
	\end{equation}
	\begin{equation}
		\begin{aligned}
			\hat{\bm{\theta}}(k+1|k+1) = \hat{\bm{\theta}}(k+1|k)+\bm{K}(k+1)\bm{\epsilon}(k+1)
		\end{aligned}
	\end{equation}
	\begin{equation}
		\begin{aligned}
			\bm{P}(k+1|k+1) = \beta(\rho) (\bm{I}-\bm{K}(k+1)\bm{\phi}(k))\bm{P}(k+1|k) \\
			(\bm{I}-\bm{K}(k+1)\bm{\phi}(k))^T +\rho^{-1}\bm{K}(k+1)\bm{Q}(k)\bm{K}^T(k+1) ,
		\end{aligned}
	\end{equation}
	where the gain $\bm{K}(k+1) \in \mathbb{R}^{s \times l}$ is 
	\begin{equation}
		\begin{aligned}
			\bm{K}(k+1) = &\bm{P}(k+1|k)\bm{\phi}^T(k) [\bm{\phi}(k)\bm{P}(k+1|k)\bm{\phi}^T(k)\\
			& +\rho^{-1}\bm{Q}(k)]^{-1},
		\end{aligned}
	\end{equation}
	the innovation $\bm{\epsilon}(k+1) \in \mathbb{R}^l$ is 
	\begin{equation} \label{innovation1}
		\begin{aligned}
			\bm{\epsilon}(k+1) = \bm{H}(k+1)-\bm{\phi}(k)\hat{\bm{\theta}}(k+1|k),
		\end{aligned}
	\end{equation}	
	and the scalar parameter $\beta(\rho)\in \mathbb{R}$ is  
	\begin{equation}
		\begin{aligned}
			\beta(\rho) = & 1+\rho-\bm{\epsilon}^T(k+1) [\rho^{-1}\bm{Q}(k) \\
			& +\bm{\phi}(k)\bm{P}(k+1|k)\bm{\phi}^T(k)]^{-1} \bm{\epsilon}(k+1).
		\end{aligned}
	\end{equation}	
\end{myTheo}

\begin{myRemark}\label{remark3}
The theory in Theorem~\ref{theo1} holds for all $ \rho \in [0, \infty)$. The optimal value of the scalar parameter $\rho$ is supposed to minimize the size of $\mathcal{E}(\bm{P}(k+1|k+1),\hat{\bm{\theta}}(k+1|k+1))$. Such an optimal value can be found by using different criteria, e.g., minimum-volume criterion, minimum-trace criterion, or minimum summation of half axes~\cite{maksarov1996state}. We illustrated how we obtain the optimal value of $\rho$ based on volume minimization in proposition~\ref{prop2}. According to proposition~\ref{prop2}, the optimal $\rho$ satisfies the following equation
\begin{equation} \label{rholamd}
	\begin{aligned}
		\sum_{i=1}^{n} \frac{\lambda_i(k) }{1+\rho \lambda_i(k) }=n\frac{\beta'(\rho)}{\beta(\rho)},
	\end{aligned}
\end{equation}
where $\lambda_i(k)$ are the eigenvalues of matrix $\bm{P}(k+1|k)\bm{\phi}(k)\bm{Q}(k)\bm{\phi}^T(k)$, and $\beta'(\rho)$ represents the derivative of $\beta(\rho)$ with respect to $\rho$, given by 
\begin{equation}
	\begin{aligned}
		&\beta'(\rho) = 1-\bm{\epsilon}^T(k+1) [\rho^{-1}\bm{Q}(k)+\bm{\phi}(k)\bm{P}(k+1|k)\bm{\phi}^T(k)]^{-1}\\ 
		& \rho^{-2}\bm{Q}(k) [\rho^{-1}\bm{Q}(k)+\bm{\phi}(k)\bm{P}(k+1|k)\bm{\phi}^T(k)]^{-1} \bm{\epsilon}(k+1).
	\end{aligned}
\end{equation}	

Note that (\ref{rholamd}) does not have a solution when 
\begin{equation}
	\begin{aligned}
		&n(1-\bm{\epsilon}^T(k+1)\bm{Q}^{-1}(k)\bm{\epsilon}^T(k+1))\\
		&-tr(\bm{P}(k+1|k)\bm{\phi}^T(k)\bm{Q}^{-1}(k)\bm{\phi}(k))>0,
	\end{aligned}
\end{equation}	
and the optimal solution is $\rho=0$ in that case.
\end{myRemark}

\begin{myRemark}
	Remark \ref{remark3} provides the optimal solution for the parameter $\rho$ used to derive the posterior ellipsoid $\mathcal{E}(\bm{P}(k+1|k+1),\hat{\bm{\theta}}(k+1|k+1))$, which can be regarded as the prior ellipsoid at the next iteration $(k+2)$
	\begin{equation}
		\begin{aligned}
				\mathcal{E}(\bm{P}(k+2|k+1),\hat{\bm{\theta}}(k+2|k+1)) \\
				= 	\mathcal{E}(\bm{P}(k+1|k+1),\hat{\bm{\theta}}(k+1|k+1)) .
		\end{aligned}
	\end{equation}
 
	 We repeat the iterative learning procedure in Theorem \ref{theo1} to gain more accurate estimation of the ellipsoid set from observations, which can reduce the uncertainties of the system during the control and as a result enhance the tracking control performance.
\end{myRemark}

\begin{myCoro}\label{coro1}
The ellipsoidal set of innovation $\bm{\epsilon}(k+1)$ described in (\ref{innovation1}) is $\mathcal{E}(\bm{P}_{\epsilon}(k+1),\bm{0})$, i.e.
\begin{equation}\label{innovationmatrix}
	\begin{aligned}
		\bm{P}_{\epsilon}(k+1) = &(q^{-1}+1)\bm{\phi}(k)\bm{P}(k)\bm{\phi}^T(k)+(q+1)\bm{Q}(k) .
	\end{aligned}
\end{equation}
\end{myCoro}
\begin{proof}
The innovation $\bm{\epsilon}(k+1)$ can be rewritten as 
\begin{equation}\label{innovation2}
	\begin{aligned}
		\bm{\epsilon}(k+1) &= \bm{\phi}(k)(\bm{\theta}(k)-\hat{\bm{\theta}}(k))+\bm{\upsilon}(k)\\
		&= \bm{\phi}(k)\tilde{\bm{\theta}}(k)+\bm{\upsilon}(k) ,
	\end{aligned}
\end{equation}

According to Theorem~\ref{theo1}, the ellipsoid for $\tilde{\bm{\theta}}(k)$ is $\mathcal{E}(\bm{P}(k),\bm{0})$, and the ellipsoid for $\bm{\phi}(k)\tilde{\bm{\theta}}(k)$ is $\mathcal{E}(\bm{\phi}(k)\bm{P}(k)\bm{\phi}^T(k),\bm{0})$. By the innovation (\ref{innovation2}), the innovation ellipsoidal set $\mathcal{E}(\bm{P}_{\epsilon}(k+1),\bm{0})$ can be calculated as 
\begin{equation}
	\begin{aligned}
		\mathcal{E}(\bm{\phi}(k)\bm{P}(k)\bm{\phi}^T(k),\bm{0}) \oplus \mathcal{E}(\bm{Q}(k),\bm{0}) \subseteq \mathcal{E}(\bm{P}_{\epsilon}(k+1),\bm{0}).
	\end{aligned}
\end{equation}

We can obtain the result in (\ref{innovationmatrix}) by applying proposition~\ref{prop1}. 
\end{proof}

\section{Robust control with one-step prediction}\label{section4}

With the learned ellipsoidal set, it is possible to derive an optimal robust control law by solving the problem ($\mathcal{P}$) in (\ref{the control objective}). However, as explained in Remark~\ref{remark1}, solving ($\mathcal{P}$) explicitly is non-trivial. To address this issue, this section derives an immediate optimal robust control law via solving a truncated one-step prediction tracking problem ($\mathcal{P}_k$), as illustrated below.
 
\begin{equation} \label{prob_k}
	\begin{aligned}
		(\mathcal{P}_k): \quad  &\min_{\bm{u}(k)} \max_{\left\lbrace \tiny\tabincell{c}{$\bm{\theta}(k) \in \mathcal{E}(\bm{P}(k), \hat{\bm{\theta}}(k))$\\$\bm{\omega}(k) \in \mathcal{E}(\bm{R}(k),\bm{0})$ }\right\rbrace   } \| \bm{y}(k+1)-\bm{y}_r(k+1) \|_{1} \\
		 \quad s.t. \quad &\bm{x}(k+1) = \bm{A}(\bm{\alpha})\bm{x}(k)+\bm{B}(\bm{\beta})\bm{u}(k)+\bm{\omega}(k),\\
		 \quad \quad \quad \quad &\bm{y}(k) = \bm{C}(k)\bm{x}(k),\\
		 \quad \quad \quad \quad &\bm{u}_{min} \leq \bm{u}(k) \leq \bm{u}_{max},\\
		 \quad \quad \quad \quad &k=1, 2,\cdots, N .
	\end{aligned}
\end{equation}

\begin{myLemma} \label{lemma1}
	The tracking error $\bm{e}(k+1)$ in the problem ($\mathcal{P}_k$) can be written as 
	\begin{equation}
		\begin{aligned}
			\bm{e}(k+1) &= \bm{y}(k+1)-\bm{y}_r(k+1) \\
			&= \bm{f}(k)+\bm{b}_0(k)\bm{u}(k)+\sum_{i=1} ^{r+s}\bm{g}_i(k)\theta_i+\bm{\upsilon}(k),
		\end{aligned}
	\end{equation}
	where $\bm{f}(k)\in \mathbb{R}^l$ is defined as
	\begin{equation}\label{paraf}
		\begin{aligned}
			\bm{f}(k) = \bm{C}(k+1)\bm{A}_0\bm{x}(k)-\bm{y}_r(k+1),
		\end{aligned}
	\end{equation}	
	$\bm{b}_0(k)\in \mathbb{R}^{l\times m}$ is defined as
	\begin{equation}\label{parab0}
		\begin{aligned}
			\bm{b}_0(k) = \bm{C}(k+1)\bm{B}_0,
		\end{aligned}
	\end{equation}	
	and	$\bm{g}_i(k)\in \mathbb{R}^{l}$ is defined as
	\begin{equation}\label{parag}
		\begin{aligned}
			\bm{g}_i(k) = \left\{ \begin{array}{rl} \bm{C}(k+1)\bm{A}_i\bm{x}(k), & i=1,2,\cdots,r\\
				\bm{C}(k+1)\bm{B}_i \bm{u}(k), &  i=r+1,\cdots,r+s. \end{array} \right.
		\end{aligned}
	\end{equation}		
\end{myLemma}

\begin{proof}
	Substituting the system dynamic (\ref{sys1}) and (\ref{sys2}) into the tracking error $e(k+1)$ yields
	\begin{equation}\label{error1}
		\begin{aligned}
			\bm{e}(k+1) =& \bm{C}(k+1)\bm{A}(\bm{\alpha})\bm{x}(k)+\bm{C}(k+1)\bm{B}(\bm{\beta})\bm{u}(k)\\
			&+ \bm{\upsilon}(k)-\bm{y}_r(k+1).
		\end{aligned}
	\end{equation}
	Combine the definition for the parameter $\bm{A}(\bm{\alpha})$ in (\ref{paraA}) and $\bm{B}(\bm{\beta})$ in (\ref{paraB}), we can obtain 
	\begin{equation}\label{error2}
		\begin{aligned}
			&\bm{e}(k+1)  = \bm{C}(k+1)\bm{A}_0\bm{x}(k)+\bm{C}(k+1)\bm{B}_0\bm{u}(k)\\
			&+\bm{C}(k+1)(\bm{A}_1\alpha_1+\bm{A}_2\alpha_2+\dots+\bm{A}_r\alpha_r)\bm{x}(k)\\
			&+\bm{C}(k+1)(\bm{B}_1\beta_1+\bm{B}_2\beta_2+\dots+\bm{B}_s\beta_s)\bm{u}(k)\\
			&+\bm{\upsilon}(k)-\bm{y}_r(k+1),\\
			& = \bm{C}(k+1)\bm{A}_0\bm{x}(k)+\bm{C}(k+1)\bm{B}_0\bm{u}(k)-\bm{y}_r(k+1)\\
			&+\bm{C}(k+1)(\bm{A}_1\bm{x}(k)\theta_1+\bm{A}_2\bm{x}(k)\theta_2+\dots+\bm{A}_r\bm{x}(k)\theta_r)\\
			&+\bm{C}(k+1)(\bm{B}_1\bm{u}(k)\theta_{r+1}+\bm{B}_2\bm{u}(k)\theta_{r+2}+\dots\\
			&+\bm{B}_s\bm{u}(k)\theta_{r+s})+\bm{\upsilon}(k).\\
		\end{aligned}
	\end{equation}
 
	The lemma is proofed with the definition in (\ref{paraf}), (\ref{parab0}) and (\ref{parag}).
\end{proof}

\begin{myLemma} \label{lemma2}
	The $t$-th element of the tracking error represented by $e^{(t)}(k+1)$ can be rewritten as
	\begin{equation}\label{tracking error}
		\begin{aligned}
			e^{(t)}(k+1) = &f^{(t)}(k)+\bm{b}_0^{(t)}(k)\bm{u}(k)+\bm{g}^{(t)}(k)\bm{\theta}\\
			&+\bm{I}_t\bm{\upsilon}(k),\quad t=1,2,\cdots,l,
		\end{aligned}
	\end{equation}
	where $f^{(t)}(k)\in \mathbb{R}$ is the $t$-th element of vector $\bm{f}(k)\in \mathbb{R}^l$, $\bm{b}_0^{(t)}(k)\in \mathbb{R}^m$ is the $t$-th row of matrix $\bm{b}_0(k)\in \mathbb{R}^{l \times m}$, and $\bm{g}^{(t)}(k)\in \mathbb{R}^s$ is defined as
	\begin{equation}
		\begin{aligned}
			\bm{g}^{(t)}(k) = [\bm{g}_1^{(t)}(k), \bm{g}_2^{(t)}(k), \cdots, \bm{g}_{r+s}^{(t)}(k)].
		\end{aligned}
	\end{equation}		
	$\bm{I}_t$ in (\ref{tracking error}) is a unit vector where the $t$-th element is $1$ and the other elements are $0$.
\end{myLemma}

\begin{myLemma}\label{lemma3}
	Given an ellipsoid in $\mathbb{R}^n$, $\mathcal{E}(\bm{P},\bm{a}) \triangleq \{ \bm{x} \in \mathbb{R}^n | (\bm{x}-\bm{a})^T \bm{P}^{-1} (\bm{x}-\bm{a}) \leq 1 \} $, where matrix $\bm{P}\in \mathbb{R}^{n \times n} $ and $\bm{P}\succ 0 $, and $\bm{a} \in \mathbb{R}^n $ is the center of the ellipsoid. For any vector $\bm{\eta} \in \mathbb{R}^n$, it holds that
	\begin{equation}\label{minE}
		\begin{aligned}
			\min_{\bm{x}\in\mathcal{E} } \bm{\eta}^T\bm{x} = \bm{\eta}^T\bm{a}-(\bm{\eta}^T\bm{P}\bm{\eta})^{1/2}, \\
		\end{aligned}
	\end{equation}		
	\begin{equation}\label{maxE}
	\begin{aligned}
		\max_{\bm{x}\in\mathcal{E} } \bm{\eta}^T\bm{x} = \bm{\eta}^T\bm{a}+(\bm{\eta}^T\bm{P}\bm{\eta})^{1/2}.
	\end{aligned}
\end{equation}		
\end{myLemma}
\begin{proof}
	We define the Lagrange function for the optimization problem $\min_{\bm{x}\in\mathcal{E} } \bm{\eta}^T\bm{x} $ in (\ref{minE}) as
	\begin{equation}
		\begin{aligned}
			L(\bm{x},\lambda) = \bm{\eta}^T\bm{x}+\lambda[(\bm{x}-\bm{a})^T \bm{P}^{-1} (\bm{x}-\bm{a})-1],
		\end{aligned}
	\end{equation}		
	and the KKT condition 
 is written as
	\begin{equation} \label{Lagra}
		\begin{aligned}
			\nabla_x L(\bm{x},\lambda)= \bm{\eta}+\lambda(2\bm{P}^{-1}\bm{x}-2\bm{P}^{-1}\bm{a}) =0 ,
		\end{aligned}
	\end{equation}		
	\begin{equation}
		\begin{aligned}
			(\bm{x}-\bm{a})^T \bm{P}^{-1} (\bm{x}-\bm{a}) -1 \leq 0,
		\end{aligned}
	\end{equation}		
	\begin{equation}
		\begin{aligned}
			\lambda \geq 0,
		\end{aligned}
	\end{equation}	
	\begin{equation} \label{cond1}
		\begin{aligned}
			\lambda [(\bm{x}-\bm{a})^T \bm{P}^{-1} (\bm{x}-\bm{a}) -1] = 0.
		\end{aligned}
	\end{equation}	
Solving (\ref{Lagra}), we obtain the solution
\begin{equation} \label{opt1}
	\begin{aligned}
		\bm{x}^*= \bm{a}-\frac{\bm{P}\bm{\eta}}{2\lambda}, \quad \lambda>0.
	\end{aligned}
\end{equation}		
Substituting (\ref{opt1}) into (\ref{cond1}) yields 
\begin{equation} \label{opt2}
	\begin{aligned}
		\lambda^* = \frac{1}{2} (\bm{\eta}^T\bm{P}\bm{\eta})^{\frac{1}{2}}. 
	\end{aligned}
\end{equation}	
Combine (\ref{opt1}) and (\ref{opt2}), and we obtain the minimization of $\bm{\eta}^T\bm{x}$
\begin{equation}
	\begin{aligned}
		\min_{\bm{x}\in\mathcal{E} } \bm{\eta}^T\bm{x} &= \bm{\eta}^T\bm{x}^*= \bm{\eta}^T \left[ \bm{a}-\frac{\bm{P}\bm{\eta}}{2\lambda^*}\right] \\
		 & = \bm{\eta}^T \left[ \bm{a}-\frac{\bm{P}\bm{\eta}}{(\bm{\eta}^T\bm{P}\bm{\eta})^{1/2}}\right]\\
		 & = \bm{\eta}^T \bm{a} - (\bm{\eta}^T\bm{P}\bm{\eta})^{1/2}.
	\end{aligned}
\end{equation}		

Similarly, we can obtain the solution of $\max_{\bm{x}\in\mathcal{E} } \bm{\eta}^T\bm{x}$ in (\ref{maxE}) by solving the problem $\min_{\bm{x}\in\mathcal{E} } -\bm{\eta}^T\bm{x} $.
\end{proof}

\begin{myTheo}\label{theo2}
Problem ($\mathcal{P}_k$) in (\ref{prob_k}) is equivalent to the following formulation
\begin{equation} \label{prob_SOCP}
	\begin{aligned}
		&(\mathcal{P}_k): \quad  \min_{\bm{u}(k)} \quad \sum_{t=1}^{l} z_t \\
		& \quad s.t. \quad  f^{(t)}(k)+\bm{b}_0^{(t)}(k)\bm{u}(k)+\bm{g}^{(t)}(k)\hat{\bm{\theta}}(k) \\
		& \quad \quad   -\left( \bm{g}^{(t)}(k)\bm{P}(k)(\bm{g}^{(t)}(k))^T\right)^{\frac{1}{2}}-\left(\bm{I}_t\bm{Q}(k)\bm{I}_t^T \right)^{\frac{1}{2}} \geq -z_t ,\\
		& \quad \quad \quad  f^{(t)}(k)+\bm{b}_0^{(t)}(k)\bm{u}(k)+\bm{g}^{(t)}(k)\hat{\bm{\theta}}(k) \\
		& \quad \quad   +\left( \bm{g}^{(t)}(k)\bm{P}(k)(\bm{g}^{(t)}(k))^T\right)^{\frac{1}{2}}+\left(\bm{I}_t\bm{Q}(k)\bm{I}_t^T \right)^{\frac{1}{2}} \leq z_t  ,\\
		& \quad \quad \quad \quad t=1,2,\cdots,l, \\
		& \quad \quad \quad \quad \bm{u}_{min} \leq \bm{u}(k) \leq \bm{u}_{max}.\\
	\end{aligned}
\end{equation}

\end{myTheo}

\begin{proof}
We firstly rewrite the problem ($\mathcal{P}_k$) as 
\begin{equation} \label{}
	\begin{aligned}
		(\mathcal{P}_k): \quad  &\min_{\bm{u}(k)} \quad \sum_{t=1}^{l} \max_{\left\lbrace \tiny\tabincell{c}{$\bm{\theta}(k) \in \mathcal{E}(\bm{P}(k), \hat{\bm{\theta}}(k))$\\$\bm{\omega}(k) \in \mathcal{E}(\bm{R}(k),\bm{0})$ }\right\rbrace   }  \|e^{(t)}(k+1)\|_1 \\
		 \quad s.t. \quad &\bm{x}(k+1) = \bm{A}(\bm{\alpha})\bm{x}(k)+\bm{B}(\bm{\beta})\bm{u}(k)+\bm{\omega}(k),\\
		 \quad \quad \quad \quad &\bm{y}(k) = \bm{C}(k)\bm{x}(k),\\
		 \quad \quad \quad \quad &\bm{u}_{min} \leq \bm{u}(k) \leq \bm{u}_{max}.\\
	\end{aligned}
\end{equation}

By introducing the auxiliary variables $z_t\geq0$, $t=1,2,\cdots,l$, we can reformulate the above problem equivalently as  
\begin{equation} \label{reformedProbSum}
	\begin{aligned}
		&(\mathcal{P}_k): \quad  \min_{\bm{u}(k)} \quad \sum_{t=1}^{l} z_t \\
		& \quad s.t. \quad \max_{\left\lbrace \tiny\tabincell{c}{$\bm{\theta}(k) \in \mathcal{E}(\bm{P}(k), \hat{\bm{\theta}}(k))$\\$\bm{\omega}(k) \in \mathcal{E}(\bm{R}(k),\bm{0})$ }\right\rbrace   }  |e^{(t)}(k+1)| \leq z_t,\\
		& \quad \quad \quad \quad \bm{x}(k+1) = \bm{A}(\bm{\alpha})\bm{x}(k)+\bm{B}(\bm{\beta})\bm{u}(k)+\bm{\omega}(k),\\
		& \quad \quad \quad \quad \bm{y}(k) = \bm{C}(k)\bm{x}(k),\\
		& \quad \quad \quad \quad \bm{u}_{min} \leq \bm{u}(k) \leq \bm{u}_{max}.\\
	\end{aligned}
\end{equation}

According to Lemma (\ref{lemma1}) and Lemma (\ref{lemma2}), we can obtain the following equation
\begin{equation} \label{temp_eq}
	\begin{aligned}
		\min_{\left\lbrace \tiny\tabincell{c}{$\bm{\theta}(k) \in \mathcal{E}(\bm{P}(k), \hat{\bm{\theta}}(k))$\\$\bm{\omega}(k) \in \mathcal{E}(\bm{R}(k),\bm{0})$ }\right\rbrace }  e^{(t)}(k+1) = f^{(t)}(k)+\bm{b}_0^{(t)}(k)\bm{u}(k) \\
		+ \min_{\left\lbrace \tiny\tabincell{c}{$\bm{\theta}(k) \in \mathcal{E}(\bm{P}(k), \hat{\bm{\theta}}(k))$\\$\bm{\omega}(k) \in \mathcal{E}(\bm{R}(k),\bm{0})$ }\right\rbrace   }\left( \bm{g}^{(t)}(k)\bm{\theta}+\bm{I}_t\bm{\upsilon}(k)\right) .
	\end{aligned}
\end{equation}

We gain the solution of (\ref{temp_eq}) by applying Lemma (\ref{lemma3}) 
\begin{equation} \label{min_eq}
	\begin{aligned}
		\min_{\left\lbrace \tiny\tabincell{c}{$\bm{\theta}(k) \in \mathcal{E}(\bm{P}(k), \hat{\bm{\theta}}(k))$\\$\bm{\omega}(k) \in \mathcal{E}(\bm{R}(k),\bm{0})$ }\right\rbrace }  e^{(t)}(k+1) = f^{(t)}(k)+\bm{b}_0^{(t)}(k)\bm{u}(k) \\
		+ \bm{g}^{(t)}(k)\hat{\bm{\theta}}(k) -\left( \bm{g}^{(t)}(k)\bm{P}(k)(\bm{g}^{(t)}(k))^T\right)^{\frac{1}{2}}\\
		-\left(\bm{I}_t\bm{Q}(k)\bm{I}_t^T \right)^{\frac{1}{2}} .
	\end{aligned}
\end{equation}

Similarly, we get the equation
\begin{equation} \label{max_eq}
	\begin{aligned}
		\max_{\left\lbrace \tiny\tabincell{c}{$\bm{\theta}(k) \in \mathcal{E}(\bm{P}(k), \hat{\bm{\theta}}(k))$\\$\bm{\omega}(k) \in \mathcal{E}(\bm{R}(k),\bm{0})$ }\right\rbrace }  e^{(t)}(k+1) = f^{(t)}(k)+\bm{b}_0^{(t)}(k)\bm{u}(k) \\
		+ \bm{g}^{(t)}(k)\hat{\bm{\theta}}(k) +\left( \bm{g}^{(t)}(k)\bm{P}(k)(\bm{g}^{(t)}(k))^T\right)^{\frac{1}{2}}\\
		+\left(\bm{I}_t\bm{Q}(k)\bm{I}_t^T \right)^{\frac{1}{2}} .
	\end{aligned}
\end{equation}

We find that the below condition 
\begin{equation} \label{}
	\begin{aligned}
		\max_{\left\lbrace \tiny\tabincell{c}{$\bm{\theta}(k) \in \mathcal{E}(\bm{P}(k), \hat{\bm{\theta}}(k))$\\$\bm{\omega}(k) \in \mathcal{E}(\bm{R}(k),\bm{0})$ }\right\rbrace   }  |e^{(t)}(k+1)| \leq z_t,\\
	\end{aligned}
\end{equation}
is equivalent to the following one 
\begin{equation} \label{condition}
	\begin{aligned}
		&\min_{\left\lbrace \tiny\tabincell{c}{$\bm{\theta}(k) \in \mathcal{E}(\bm{P}(k), \hat{\bm{\theta}}(k))$\\$\bm{\omega}(k) \in \mathcal{E}(\bm{R}(k),\bm{0})$ }\right\rbrace   }  e^{(t)}(k+1) \geq -z_t,\\
		&\max_{\left\lbrace \tiny\tabincell{c}{$\bm{\theta}(k) \in \mathcal{E}(\bm{P}(k), \hat{\bm{\theta}}(k))$\\$\bm{\omega}(k) \in \mathcal{E}(\bm{R}(k),\bm{0})$ }\right\rbrace   }  e^{(t)}(k+1) \leq z_t.\\
	\end{aligned}
\end{equation}
Substituting (\ref{min_eq}) and (\ref{max_eq}) into the condition (\ref{condition}) yields 
\begin{equation} \label{condition2}
	\begin{aligned}
		& f^{(t)}(k)+\bm{b}_0^{(t)}(k)\bm{u}(k)+\bm{g}^{(t)}(k)\hat{\bm{\theta}}(k) \\
		& -\left( \bm{g}^{(t)}(k)\bm{P}(k)(\bm{g}^{(t)}(k))^T\right)^{\frac{1}{2}}-\left(\bm{I}_t\bm{Q}(k)\bm{I}_t^T \right)^{\frac{1}{2}} \geq -z_t ,\\
		& f^{(t)}(k)+\bm{b}_0^{(t)}(k)\bm{u}(k)+\bm{g}^{(t)}(k)\hat{\bm{\theta}}(k) \\
		& +\left( \bm{g}^{(t)}(k)\bm{P}(k)(\bm{g}^{(t)}(k))^T\right)^{\frac{1}{2}}+\left(\bm{I}_t\bm{Q}(k)\bm{I}_t^T \right)^{\frac{1}{2}} \leq z_t  .\\
	\end{aligned}
\end{equation}
Combing (\ref{condition2}) and (\ref{reformedProbSum}), we then obtain the result in Theorem~\ref{theo2}.
\end{proof}
\begin{myRemark}
	After the reformulation of the problem ($\mathcal{P}_k$) in Theorem~\ref{theo2}, we translate ($\mathcal{P}_k$) into a typical second-order cone programming (SOCP) problem, which can be efficiently solved, e.g., by the interior point method~\cite{alizadeh2003second}.
\end{myRemark}

\section{Active learning for robust control}\label{section5}

In this section, we enhance the quality of the learned ellipsoidal set by mathematically maximizing the ellipsoidal set volume. The maximum set volume leads to enriched information fed to the set learning procedure, resulting in an extended learning capacity and enhanced learning accuracy for the ellipsoidal set. The increased set volume nevertheless renders the control system to handle higher levels of uncertainty during the control. To address this dilemma between learning and control, we adopt the bi-criterion approach and gain an efficient balance, and ultimately derive the optimal robust control law with fully-fledged active learning. We detail this effort as follows.

According to the Theorem~\ref{theo2} in section~\ref{section4}, the tracking performance highly depends on the quality of the learned ellipsoidal set for the uncertainties. One efficient method to improve the learning quality and speed up the learning process of the ellipsoidal set is enriching the information used for learning. From Theorem~\ref{theo1}, it can be observed that the innovation (\ref{innovation1}) contains new information $\bm{H}(k+1)$ that is needed in the learning. Hence, it is natural to enlarge the innovation for information enrichment. 

According to Corollary (\ref{coro1}), the ellipsoidal set of innovation $\bm{\epsilon}(k+1)$ is $\mathcal{E}(\bm{P}_{\epsilon}(k+1),0)$, where the matrix $\bm{P}_{\epsilon}(k+1)$ describes the shape of the ellipsoid. We can increase the innovation by maximizing the trace of the shape matrix $\bm{P}_{\epsilon}(k+1)$ for the information enrichment. To achieve this, the following optimization objective is added to the control.
\begin{equation} \label{prob_a}
	\begin{aligned}
		& (\mathcal{P}_a): \quad  \max_{\bm{u}(k)} \quad \text{tr}(\bm{P}_{\epsilon}(k+1)) .\\
	\end{aligned}
\end{equation}
Solving the problem ($\mathcal{P}_a$) might lead to unbounded large control values that are inapplicable; therefore, there is a need to design a constraint to restrict the control values. We provide the solution for this issue and explain how the optimization objective ($\mathcal{P}_a$) connects the derivation of our optimal control law in the following.

Note that Theorem~\ref{theo2} results in a robust control law $\bm{u}_r(k)$ that considers the worst-case performance within ellipsoidal sets $\mathcal{E}(\bm{P}(k), \hat{\bm{\theta}}(k))$ and $\mathcal{E}(\bm{R}(k),\bm{0})$. Therefore, the resulted control law is a comparatively conservative or cautious control strategy. I.e., the resulted control law is limited and does not have the active learning or exploring ability to learn the ellipsoidal set in a wide potential range. With that said, the objective ($\mathcal{P}_a$) provides the opportunity of deriving an active learning featured control law $\bm{u}_a(k)$ that can excite the system and accelerate the uncertainty learning process. However, such excitation results in large control values that brings large overshoot, and vibrates or even jeopardizes the system control. In that sense, the optimal one-step predictive robust control $\bm{u}_r(k)$ and the active uncertainty learning $\bm{u}_a(k)$ are conflicting. 

To get a reasonable compromise for this conflict, we design a constraint $\Omega(k)$ that is symmetrically located around the robust control law $\bm{u}_r(k)$, and then the optimal active learning featured control law will depend on the size of $\Omega(k)$. The domain $\Omega(k)$ is defined as
\begin{equation} \label{omega_domain}
	\begin{aligned}
		\Omega(k) = \{ \bm{u}(k)\in \mathbb{R}^m | &(\bm{u}(k)-\bm{u}_r(k))^T(\text{tr}(\bm{P}(k)) \bm{M}_a)^{-1}\\
		&(\bm{u}(k)-\bm{u}_r(k)) \leq 1 \} ,\\
	\end{aligned}
\end{equation}
where $\bm{M}_a\in \mathbb{R}^{m\times m}$ is a positive definite weighting matrix, $\bm{P}(k)$ is the matrix defining the shape of ellipsoid for unknown parameter $\bm{\theta}$, $\text{tr}(\cdot)$ represents the matrix trace. Under this definition, the size of $\Omega(k)$ depends on the accuracy of estimation, or the uncertainty of parameter, which is characterized by matrix $\bm{P}(k)$ illustrated in Theorem~\ref{theo1}. From (\ref{omega_domain}) we can see the domain $\Omega(k)$ is an ellipsoid centered at $\bm{u}_r(k)$, which can be written as $\mathcal{E}(\text{tr}(\bm{P}(k)) \bm{M}_a, \bm{u}_r(k))$.

With the designed domain $\Omega(k)$, we can derive the uncertainty active learning featured adaptive robust control law by solving the following optimization problem 
\begin{equation} \label{activeLProb}
	\begin{aligned}
		&(\mathcal{P}_a): \quad  \max_{\bm{u}(k) \in \Omega(k) } \quad \text{tr}(\bm{P}_{\epsilon}(k+1)) \\
		& \quad s.t. \quad \Omega(k) = \mathcal{E}(\text{tr}(\bm{P}(k)) \bm{M}_a, \bm{u}_r(k)),\\
	\end{aligned}
\end{equation}

\begin{myTheo} \label{theo3}
Problem ($\mathcal{P}_a$) in (\ref{activeLProb}) is equivalent to the following formulation
\begin{equation} \label{activeLProbReform}
	\begin{aligned}
		&(\mathcal{P}_a):  \max_{\bm{u}(k) \in \Omega(k) } \quad   \bm{u}^T \bm{P}_W \bm{u} + \bm{x}^T \bm{P}_{VW} \bm{u}  \\
		& \quad s.t. \quad \Omega(k) = \mathcal{E}(\text{tr}(\bm{P}(k)) \bm{M}_a, \bm{u}_r(k)),\\
	\end{aligned}
\end{equation}
where the matrix $\bm{P}_W$ and $\bm{P}_{VW}$ are defined as 
\begin{equation} \label{matrixPtrace}
	\begin{aligned}
		\bm{P}_{W}  = \sum_{t=1}^{l} \bm{P}_W^t, \quad \bm{P}_{VW}  = \sum_{t=1}^{l} \bm{P}_{VW}^t,
	\end{aligned}
\end{equation}
and the matrix $\bm{P}_W^t$ and $\bm{P}_{VW}^t$ in (\ref{matrixPtrace}) are defined as 
\begin{equation} \label{}
	\begin{aligned}
		\bm{P}_W^t & = (\bm{W}_1^t)^T\bm{P}_{r+1,r+1}\bm{W}_1^t+\cdots+(\bm{W}_s^t)^T\bm{P}_{r+s,r+1}\bm{W}_1^t\\
		&+(\bm{W}_1^t)^T\bm{P}_{r+1,r+2}\bm{W}_2^t+\cdots+(\bm{W}_s^t)^T\bm{P}_{r+s,r+2}\bm{W}_2^t \\
		&+\cdots\\
		&+(\bm{W}_1^t)^T\bm{P}_{r+1,r+s}\bm{W}_s^t+\cdots+(\bm{W}_s^t)^T\bm{P}_{r+s,r+s}\bm{W}_s^t,\\
	\end{aligned}
\end{equation}	
and
\begin{equation} \label{}
	\begin{aligned}
		\bm{P}_{VW}^t & = (\bm{V}_1^t)^T\bm{P}_{1,r+1}\bm{W}_1^t+\cdots+(\bm{V}_r^t)^T\bm{P}_{r,r+1}\bm{W}_1^t\\
		&+(\bm{V}_1^t)^T\bm{P}_{1,r+2}\bm{W}_2^t+\cdots+(\bm{V}_r^t)^T\bm{P}_{r,r+2}\bm{W}_2^t \\
		&+\cdots\\
		&+(\bm{V}_1^t)^T\bm{P}_{1,r+s}\bm{W}_s^t+\cdots+(\bm{V}_r^t)^T\bm{P}_{r,r+s}\bm{W}_s^t,\\
	\end{aligned}
\end{equation}	
The matrix $\bm{V}_i \in \mathbb{R}^{l\times n}$ and $\bm{W}_j \in \mathbb{R}^{l\times m}$ are defined as
\begin{equation} \label{matrix_V}
	\begin{aligned}
		\bm{V}_i = \bm{C}\bm{A}_i , \quad i= 1,2,\cdots, r,
	\end{aligned}
\end{equation}	
\begin{equation} \label{matrix_W}
	\begin{aligned}
		\bm{W}_j = \bm{C}\bm{B}_j , \quad j= 1,2,\cdots, s.
	\end{aligned}
\end{equation}	
and the $t$-th row of matrix is represented as $\bm{V}_i^t$ and $\bm{W}_j^t$, $t=1,2,\dots,l$.
\end{myTheo}
\begin{proof}
Substituting $\bm{P}_{\epsilon}(k+1)$ represented by (\ref{innovationmatrix}) into the cost function $\text{tr}(\bm{P}_{\epsilon}(k+1))$ and adopting $\text{tr}(\bm{A}+\bm{B})=\text{tr}(\bm{A})+\text{tr}(\bm{B})$ yields 
\begin{equation} \label{innov_reform1}
	\begin{aligned}
		\text{tr}(\bm{P}_{\epsilon}(k+1)) = &(q^{-1}+1)\text{tr}(\bm{\phi}(k)\bm{P}(k)\bm{\phi}^T(k))\\
		&+(q+1)\text{tr}(\bm{Q}(k)).
	\end{aligned}
\end{equation}	
To make the control signal $\bm{u}$ visible in the cost function, we take apart the matrix $\bm{P}(k)$ and $\bm{\phi}(k)$ as
\begin{equation} \label{partMatrix}
	\begin{aligned}
		\bm{P}(k) =\left[ \begin{array}{cc} \bm{P}_{xx} & \bm{P}_{xu} \\ \bm{P}_{ux} & \bm{P}_{uu} \end{array} \right], \quad \bm{\phi}(k) =\left[ \begin{array}{cc} \bm{\phi}_{x} & \bm{\phi}_{u}   \end{array} \right].
	\end{aligned}
\end{equation}	
With the partitioned matrices (\ref{partMatrix}), we obtain the following equation
\begin{equation} \label{}
	\begin{aligned}
		&\text{tr}(\bm{\phi}(k)\bm{P}(k)\bm{\phi}^T(k)) \\
		& = \text{tr} \left( \left[ \begin{array}{cc} \bm{\phi}_{x} & \bm{\phi}_{u} \end{array} \right] \left[ \begin{array}{cc} \bm{P}_{xx} & \bm{P}_{xu} \\ \bm{P}_{ux} & \bm{P}_{uu} \end{array} \right] \left[ \begin{array}{cc} \bm{\phi}_{x} & \bm{\phi}_{u} \end{array} \right]^T \right) \\
		& =  \text{tr} \left(\bm{\phi}_{x}\bm{P}_{xx}\bm{\phi}_{x}^T\right)  + \text{tr} \left(\bm{\phi}_{x}\bm{P}_{xu}\bm{\phi}_{u}^T\right)  + \text{tr} \left(\bm{\phi}_{u}\bm{P}_{uu}\bm{\phi}_{u}^T\right),
	\end{aligned}
\end{equation}	
where the matrix $\bm{P}_{uu}$ is 
\begin{equation} \label{matrixPuu}
	\begin{aligned}
		\bm{P}_{uu} = \left[ \begin{array}{cccc} \bm{P}_{r+1,r+1} & \bm{P}_{r+1,r+2} & \cdots & \bm{P}_{r+1,r+s} \\ \bm{P}_{r+2,r+1} & \bm{P}_{r+2,r+2} & \cdots & \bm{P}_{r+2,r+s} \\ \vdots & \vdots & \cdots & \vdots \\  \bm{P}_{r+s,r+1} & \bm{P}_{r+s,r+2} & \cdots & \bm{P}_{r+s,r+s} \end{array}  \right]  ,
	\end{aligned}
\end{equation}	
and the matrix $\bm{P}_{xu}$ is 
\begin{equation} \label{matrixPxu}
	\begin{aligned}
		\bm{P}_{xu} = \left[ \begin{array}{cccc} \bm{P}_{1,r+1} & \bm{P}_{1,r+2} & \cdots & \bm{P}_{1,r+s} \\ \bm{P}_{2,r+1} & \bm{P}_{2,r+2} & \cdots & \bm{P}_{2,r+s} \\ \vdots & \vdots & \cdots & \vdots \\  \bm{P}_{r,r+1} & \bm{P}_{r,r+2} & \cdots & \bm{P}_{r,r+s} \end{array}  \right]  ,
	\end{aligned}
\end{equation}	
and with the defined matrix $\bm{V}_i$ (\ref{matrix_V}) and $\bm{W}_j$ (\ref{matrix_W}), the matrix $\bm{\phi}_{x}$ and $\bm{\phi}_{u}$ are shown as
\begin{equation} \label{}
	\begin{aligned}
		\bm{\phi}_{x}& = [\bm{C}\bm{A}_1\bm{x}(k), \bm{C}\bm{A}_2\bm{x}(k), \cdots, \bm{C}\bm{A}_r\bm{x}(k)]\\
		& = [\bm{V}_1\bm{x}(k), \bm{V}_2\bm{x}(k), \cdots, \bm{V}_r\bm{x}(k)],\\
	\end{aligned}
\end{equation}	
\begin{equation} \label{phi_u}
	\begin{aligned}
		\bm{\phi}_{u}& = [\bm{C}\bm{B}_1\bm{u}(k), \bm{C}\bm{B}_2\bm{u}(k), \cdots, \bm{C}\bm{B}_s\bm{u}(k)]\\
		& = [\bm{W}_1\bm{u}(k), \bm{W}_2\bm{u}(k), \cdots, \bm{W}_s\bm{u}(k)].\\
	\end{aligned}
\end{equation}	 
With the definition (\ref{matrixPuu}) and (\ref{phi_u}), the trace of $\bm{\phi}_{u}\bm{P}_{uu}\bm{\phi}_{u}^T$ can be calculated as
\begin{equation} \label{trace_uu}
	\begin{aligned}
		&\text{tr}(\bm{\phi}_{u}\bm{P}_{uu}\bm{\phi}_{u}^T) \\
		= & \text{tr} \left( \left[ \begin{array}{cccc} \bm{u}^T(\bm{W}_1^1)^T & \bm{u}^T(\bm{W}_2^1)^T & \cdots & \bm{u}^T(\bm{W}_s^1)^T \\ \bm{u}^T(\bm{W}_1^2)^T & \bm{u}^T(\bm{W}_2^2)^T & \cdots & \bm{u}^T(\bm{W}_s^2)^T \\ \vdots & \vdots & \cdots & \vdots \\  \bm{u}^T(\bm{W}_1^l)^T & \bm{u}^T(\bm{W}_2^l)^T & \cdots & \bm{u}^T(\bm{W}_s^l)^T \end{array}  \right] \right. \\
		& 	\left[ \begin{array}{cccc} \bm{P}_{r+1,r+1} & \bm{P}_{r+1,r+2} & \cdots & \bm{P}_{r+1,r+s} \\ \bm{P}_{r+2,r+1} & \bm{P}_{r+2,r+2} & \cdots & \bm{P}_{r+2,r+s} \\ \vdots & \vdots & \cdots & \vdots \\  \bm{P}_{r+s,r+1} & \bm{P}_{r+s,r+2} & \cdots & \bm{P}_{r+s,r+s} \end{array}  \right] \\
		&\left. \left[ \begin{array}{cccc} \bm{W}_1^1\bm{u} & \bm{W}_1^2\bm{u} & \cdots & \bm{W}_1^l\bm{u} \\ \bm{W}_2^1\bm{u} & \bm{W}_2^2\bm{u} & \cdots & \bm{W}_2^l\bm{u} \\ \vdots & \vdots & \cdots & \vdots \\  \bm{W}_s^1\bm{u} & \bm{W}_s^2\bm{u} & \cdots & \bm{W}_s^l\bm{u} \end{array}  \right] \right) \\
		& =  \bm{u}^T(\bm{P}_W^1+\bm{P}_W^2+\cdots+\bm{P}_W^t)\bm{u}\\
		& = \bm{u}^T \left( \sum_{t=1}^{l} \bm{P}_W^t \right) \bm{u} = \bm{u}^T \bm{P}_W \bm{u}.
	\end{aligned}
\end{equation}	

The derivation of $\text{tr}(\bm{\phi}_{x}\bm{P}_{xu}\bm{\phi}_{u}^T)$ is the same as (\ref{trace_uu}) and the result is shown below
\begin{equation} \label{}
	\begin{aligned}
		&\text{tr}(\bm{\phi}_{x}\bm{P}_{xu}\bm{\phi}_{u}^T) \\
		= & \text{tr} \left( \left[ \begin{array}{cccc} \bm{x}^T(\bm{V}_1^1)^T & \bm{x}^T(\bm{V}_2^1)^T & \cdots & \bm{x}^T(\bm{V}_r^1)^T \\ \bm{x}^T(\bm{V}_1^2)^T & \bm{x}^T(\bm{V}_2^2)^T & \cdots & \bm{x}^T(\bm{V}_r^2)^T \\ \vdots & \vdots & \cdots & \vdots \\  \bm{x}^T(\bm{V}_1^l)^T & \bm{x}^T(\bm{V}_2^l)^T & \cdots & \bm{x}^T(\bm{V}_r^l)^T \end{array}  \right] \right. \\
		& \left[ \begin{array}{cccc} \bm{P}_{1,r+1} & \bm{P}_{1,r+2} & \cdots & \bm{P}_{1,r+s} \\ \bm{P}_{2,r+1} & \bm{P}_{2,r+2} & \cdots & \bm{P}_{2,r+s} \\ \vdots & \vdots & \cdots & \vdots \\  \bm{P}_{r,r+1} & \bm{P}_{r,r+2} & \cdots & \bm{P}_{r,r+s} \end{array}  \right] \\
		&\left. \left[ \begin{array}{cccc} \bm{W}_1^1\bm{u} & \bm{W}_1^2\bm{u} & \cdots & \bm{W}_1^l\bm{u} \\ \bm{W}_2^1\bm{u} & \bm{W}_2^2\bm{u} & \cdots & \bm{W}_2^l\bm{u} \\ \vdots & \vdots & \cdots & \vdots \\  \bm{W}_s^1\bm{u} & \bm{W}_s^2\bm{u} & \cdots & \bm{W}_s^l\bm{u} \end{array}  \right] \right) \\
		& =  \bm{x}^T(\bm{P}_{VW}^1+\bm{P}_{VW}^2+\cdots+\bm{P}_{VW}^l)\bm{u}\\
		& = \bm{x}^T \left( \sum_{t=1}^{l} \bm{P}_{VW}^t \right) \bm{u}=\bm{x}^T \bm{P}_{VW} \bm{u}.
	\end{aligned}
\end{equation}	
Therefore the cost function $\text{tr}(\bm{P}_{\epsilon}(k+1))$ in (\ref{innov_reform1}) can be rewritten as
\begin{equation} \label{}
	\begin{aligned}
		&\text{tr}(\bm{P}_{\epsilon}(k+1)) \\
		= &(q^{-1}+1)\left( \bm{u}^T \bm{P}_W \bm{u} + \bm{x}^T \bm{P}_{VW} \bm{u} + c_1 \right) +(q+1)\text{tr}(\bm{Q}(k))	\\
		= &(q^{-1}+1)\left( \bm{u}^T \bm{P}_W \bm{u} + \bm{x}^T \bm{P}_{VW} \bm{u} \right) + C_{const}  
	\end{aligned}
\end{equation}	
where $c_1$ and $ C_{const}$ represent constant values and have nothing to do with $ \bm{u}(k)$. According to proposition \ref{prop1}, the parameter $q$ is larger than $0$, so that $q^{-1}+1$ is larger than $1$. Therefore it will not change the derived result by omitting $q^{-1}+1$.
\end{proof}
\begin{myRemark}
	After the reformulation of the problem ($\mathcal{P}_a$) in Theorem~\ref{theo3}, we translate ($\mathcal{P}_k$) into a typical quadratic constrained quadratic programming (QCQP) problem, which can be efficiently solved, e.g., by the interior point method~\cite{bertsekas2015convex}.
\end{myRemark}

Table~\ref{table1} summarizes implementation procedures of our proposed adaptive robust tracking control with active learning.

\begin{table}[htp]
	\centering
	\caption{The procedures of adaptive robust tracking control with active learning.}
	\resizebox{10cm}{!}
	{\begin{tabular}{l}
			\toprule
			\textbf{Initialization:}\\
			Initialize the ellipsoid of the uncertain parameter $\mathcal{E}(\bm{P}(0),\bm{\theta}(0))$, \\
			system output $\bm{y}(0) $, the user-defined coefficient $\lambda>0$.\\
			\midrule
			\textbf{Computation:}\\
			(1) For given $\mathcal{E}(\bm{P}(k),\bm{\theta}(k))$, solve the problem ($\mathcal{P}_k$) by \\
			\quad \quad Theorem \ref{theo2}, which yields the robust control law $\bm{u}_r$.\\
			(2) With the derived $\bm{u}_r$, solve the problem ($\mathcal{P}_a$) by \\
			\quad \quad Theorem \ref{theo3} and yields the active learning featured\\
			\quad \quad control law $\bm{u}_a$. \\
			(3) Apply the control law $\bm{u}_a$ to the system.\\
			(4) Measure the new system output $\bm{y}(k+1)$, and calculate \\
			\quad \quad $\bm{H}(k+1)$ and $\bm{\phi}(k)$. \\
			(5) Use the estimation procedure in Theorem \ref{theo1} to update a \\
			\quad \quad new ellipsoid $\mathcal{E}(\bm{P}(k+1),\bm{\theta}(k+1))$;\\
			(6) Go back to step (1);  \\
			\bottomrule
	\end{tabular}}
	\label{table1}
\end{table}

\section{Simulations and Results}\label{section6}

In this section, we test the performance of the proposed adaptive robust control by three simulations. We use diversified settings when conducting the simulations. The simulation in section~\ref{subsection1} considers the tracking control for a single-input single-output (SISO) model with two uncertain parameters, and section~\ref{subsection2} considers the same system with three uncertain parameters. Section~\ref{subsection3} conducts the tracking control for a transport aircraft model with its longitudinal dynamic, which is a multiple-input multiple-output (MIMO) system with four uncertain parameters. We analyze the simulation results regarding the tracking control performance and the ellipsoidal set learning process, and demonstrate the advantage of our approach via comparison with the set learning strategy without active learning.

\subsection{Simulation on a SISO system with two-dimensional uncertainty}\label{subsection1}

This section simulates robust tracking for the system described in (\ref{sys1}), whose system dynamics are described as
\begin{equation}
	\begin{aligned}
		\bm{A} = \bm{A}_0 = \left[ \begin{array}{ccc} 0 & 1 & 0 \\ 0 & 0 & 1 \\ -0.3 & 0.4 & 0.2 \end{array} \right],       \\
	\end{aligned} 
\end{equation}
and 
\begin{equation}
	\begin{aligned}
		\bm{B}(\bm{\beta})&=\bm{B}_0+\beta_1\bm{B}_1+\beta_2\bm{B}_2\\
		&=\left[ \begin{array}{c} -0.8 \\ 0.7 \\ -0.5 \end{array} \right] + \beta_1 \left[ \begin{array}{c} 0.3 \\ -0.6 \\ 0.3 \end{array} \right]+\beta_2 \left[ \begin{array}{c} -0.5 \\ 0.5 \\ -0.3 \end{array} \right],\\
	\end{aligned} 
\end{equation}
where the parameters $\beta_1$ and $\beta_2$ are unknown. The observation matrix in (\ref{sys2}) is instantiated as
\begin{equation}
	\begin{aligned}
		\bm{C}(k) = \left[ \begin{array}{ccc} 0.5 & 0.8 & 1.1  \end{array} \right]. 
	\end{aligned} 
\end{equation}
The following matrix 
\begin{equation}
	\begin{aligned}
		\bm{R}(k) = \left[ \begin{array}{ccc} 0.5 & 0 & 0 \\ 0 & 0.5 & 0 \\ 0 & 0 & 0.5 \end{array} \right]
	\end{aligned} 
\end{equation}
defines the shape of ellipsoid (\ref{disturbellip}) confining the disturbance $\bm{\omega}(k)$. The target system output trajectory is set as a triangle wave derived by the function 
\begin{equation}
	\begin{aligned}
		y_r(k) = \left\lbrace  \begin{array}{rl}
			5/23k, & k=1,2,\cdots,23 \\ 5-5/23(k-24), & k=24,25,\cdots,46. 
		\end{array} \right. 
	\end{aligned} 
\end{equation}

With the above system settings, we first carry out a one-time simulation to show the process of vanilla ellipsoidal set learning and the designed active learning, along with the comparison of trajectory tracking performance. In this simulation, the true value for the unknown parameter $\bm{\theta}$ is set as $\bm{\theta}^{\ast} = [0.2, 0.1]^T$, the initial state is set as $x(:,1) = [1, 1, 1]^T$, the control signal is confined in the constraint $-25 \leq u(k) \leq 25 $. The initial uncertainty ellipsoidal set is $\mathcal{E}(\bm{P}(1),\bm{\theta}(1))$, where the matrix and center are respectively set as
\begin{equation}
	\begin{aligned}
		\bm{P}(1) = \left[ \begin{array}{cc} 4 & 1 \\ 1 & 2 \end{array} \right], \quad \bm{\theta}(1) = \left[ \begin{array}{c} 0.8 \\ 0.7  \end{array} \right].
	\end{aligned} 
\end{equation}
The active learning coefficient for our proposed method is set as $M_A = 0.8$.

\begin{figure}[htbp]
	\centering
	\includegraphics[width=0.7\textwidth]{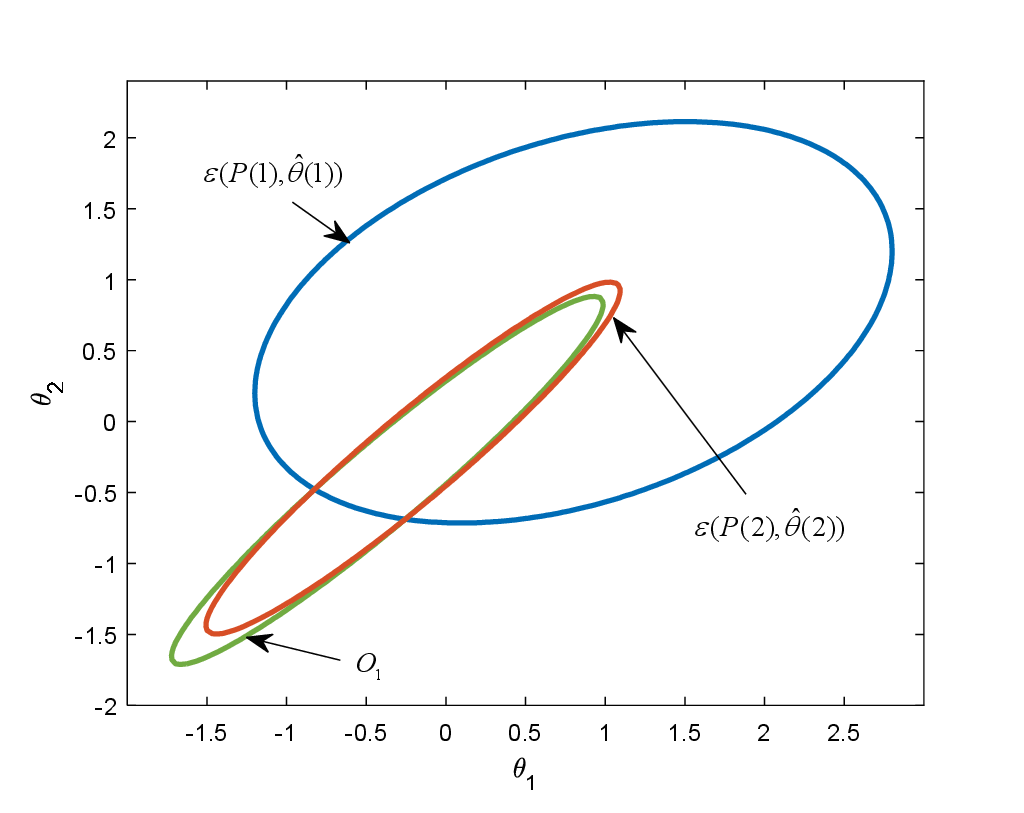}\\
	\caption{Vanilla ellipsoidal set learning at the $1$-th iteration, $\mathcal{E}(\bm{P}(2),\hat{\bm{\theta}}(2)) \supseteq \mathcal{E}(\bm{P}(1),\hat{\bm{\theta}}(1)) \cap O_1$. }\label{ellipsoidinter}
\end{figure}

Fig.~\ref{ellipsoidinter} shows result of the vanilla ellipsoidal set learning at the $1$-th iteration. The observation $O_1=\mathcal{E}(\bm{\phi}^{-1}(1)\bm{Q}(1)\bm{\phi}^{-T}(1), \bm{\phi}^{-1}(1)\bm{H}(2))$ is used to update the ellipsoidal set $\mathcal{E}(\bm{P}(1),\hat{\bm{\theta}}(1))$. $\mathcal{E}(\bm{P}(2),\hat{\bm{\theta}}(2))$ is the learned set of a minimal ellipsoid bounding the intersection of $\mathcal{E}(\bm{P}(1),\hat{\bm{\theta}}(1))$ and $O_1$, written as $\mathcal{E}(\bm{P}(2),\hat{\bm{\theta}}(2)) \supseteq \mathcal{E}(\bm{P}(1),\hat{\bm{\theta}}(1)) \cap O_1$.

\begin{figure}[H]
	\centering
	\includegraphics[width=0.7\textwidth]{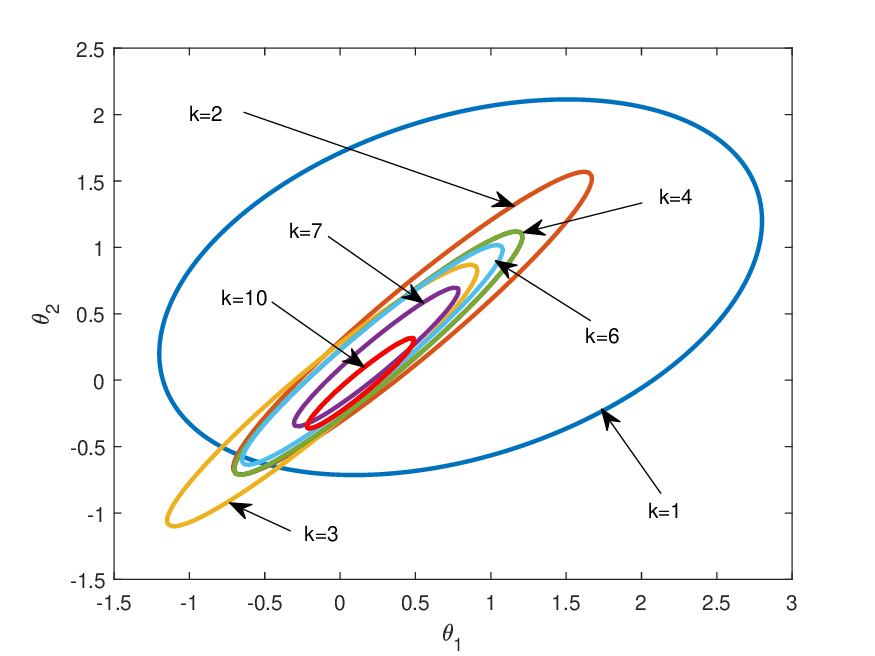}\\
	\caption{The evolution of $\mathcal{E}(\bm{P}(k),\bm{\theta}(k))$ using vanilla set learning.}\label{ellipsoid1}
\end{figure}

\begin{figure}[H]
	\centering
	\includegraphics[width=0.7\textwidth]{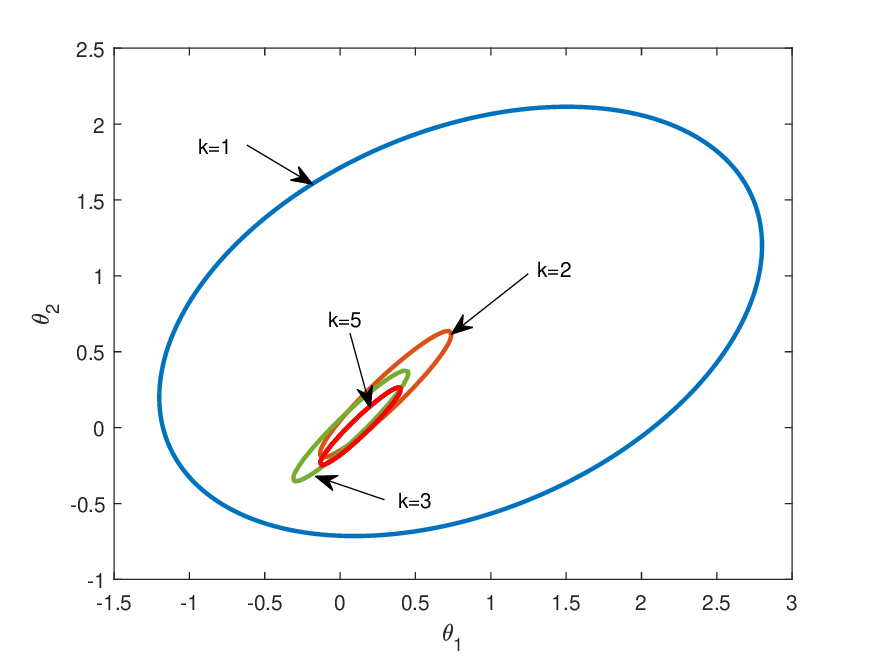}\\
	\caption{The evolution of $\mathcal{E}(\bm{P}(k),\bm{\theta}(k))$ using active learning.}\label{ellipsoid2}
\end{figure}

Fig.~\ref{ellipsoid1} and Fig.~\ref{ellipsoid2} show the evolution of the learned uncertainty ellipsoidal set $\mathcal{E}(\bm{P}(k),\bm{\theta}(k))$ from $1$-th to $10$-th iteration, for the vanilla set learning and the proposed active learning, respectively. The results manifest that the ellipsoidal sets $\mathcal{E}(\bm{P}(k),\bm{\theta}(k))$ under both of the two learning approaches shrink over time, implying reduced uncertainties over time. In comparison, the ellipsoidal set by active learning converges faster than the vanilla set learning. The ellipsoid in Fig.~\ref{ellipsoid2} under active learning converges after $5$ iterations, while that in Fig.~\ref{ellipsoid1} converges after $10$ iterations.

\begin{figure}[htbp]
	\centering
	\includegraphics[width=0.7\textwidth]{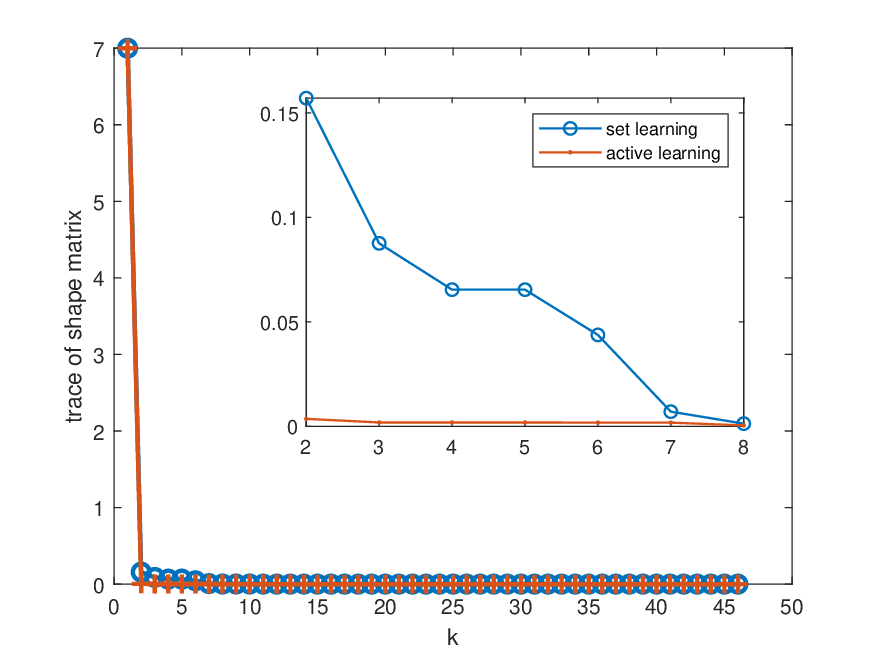}\\
	\caption{The trace of shape matrix for uncertainty ellipsoidal set with set learning and active learning.}\label{volumconverge}
\end{figure}

The trace of the shape matrix for ellipsoid can be used to quantify the uncertainty. Fig.~\ref{volumconverge} plots the trace of the shape matrix over the learning iterations for visualization. The plot shows that the shape matrix trace under both vanilla set learning and active learning is reduced closely to $0$ over time. The trace by active learning converges to $0$ after $4$ iterations, compared with $8$ iterations by vanilla set learning. This implies the active learning pushes the system to learn faster compared with vanilla set learning.

\begin{figure}[htbp]
	\centering
	\includegraphics[width=0.7\textwidth]{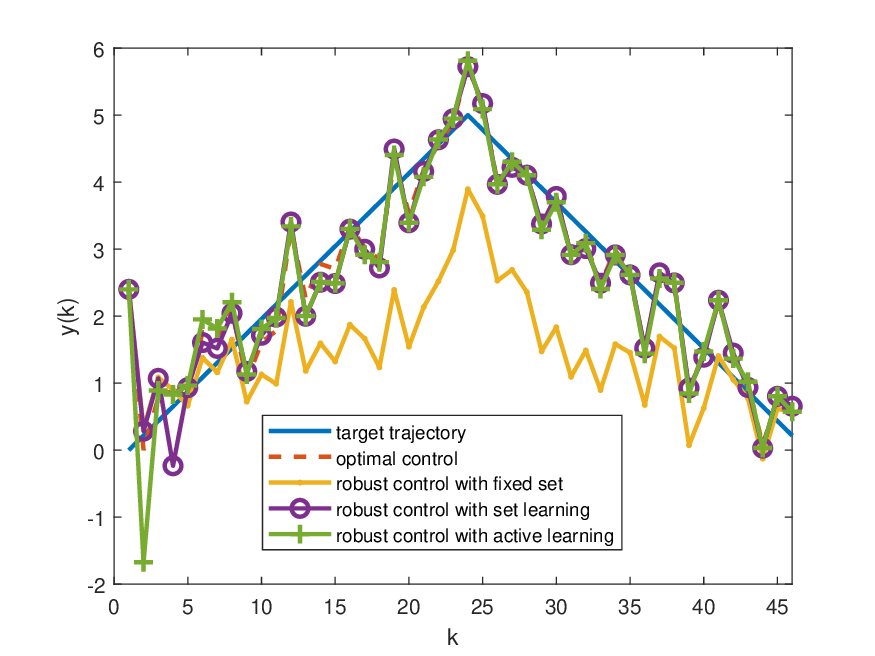}\\
	\caption{The system output during the target trajectory tracking by optimal control, robust control with fixed set, robust control with (vanilla) set learning, and robust control with active learning.}\label{control1}
\end{figure}

Fig.~\ref{control1} shows the performance of target trajectory tracking under four control methods: optimal control, robust control with fixed set, robust control with vanilla set learning, and the designed robust control with active learning. We derive and use the optimal control as an ideal benchmark that knows the true value of the parameter vector $\bm{\theta}$ during the control law derivation, which is completely unknown in the other three control strategies. We observe from Fig.~\ref{control1} that, compared with robust control with fixed uncertainty set, robust control with either vanilla or active set learning is more approximate to optimal control. Robust control with set learning mechanisms provides more accurate trajectory tracking, except for the start-up period. I.e., We observe active learning brings a larger overshoot than vanilla set learning at the beginning of the control. Such an overshoot is due to the excitation by the active learning mechanism, which is designed to stimulate the system to explore more information to feed the uncertainty learning.

\begin{table}[H]
	\caption{Control performance comparison between robust control with fixed set, vanilla set learning, and active learning.}
	\centering
	\resizebox{10cm}{!}
	{\begin{tabular}{ccccccccc}
			\toprule
			T &  $\bar{e}_1$ & $\bar{e}_2$ &  $\bar{e}_3$ & ratio1 & ratio2 & ratio3\\ 
			\midrule
			2  & 1.4294 & 1.4194 & 2.0296 & 0.70\%  & -29.57\% & -30.06\%\\ 
			4  & 0.9311 & 0.9308 & 1.2796 & 0.03\%  & -27.23\% & -27.26\%\\
			6  & 0.7888 & 0.7751 & 1.0078 & 1.77\%  & -21.73\% & -23.10\%\\
			8  & 0.7484 & 0.6940 & 0.8633 & 7.83\%  & -13.31\% & -19.61\%\\
			10 & 0.7477 & 0.6424 & 0.7721 & 16.39\% & -3.16\%  & -16.80\%\\
			\midrule
			15 & 1.0090 & 0.4359 & 0.4180 & 131.50\%& 141.41\% & 4.28\%\\
			20 & 1.2404 & 0.4331 & 0.4294 & 186.38\%& 188.86\% & 0.87\% \\
			25 & 1.4322 & 0.4279 & 0.4228 & 234.74\%& 238.76\% & 1.20\% \\
			30 & 1.4976 & 0.4216 & 0.4161 & 255.25\%& 259.92\% & 1.32\% \\
			45 & 1.2431 & 0.4117 & 0.4088 & 201.91\%& 204.04\% & 0.71\% \\
			\bottomrule
	\end{tabular}}
	\label{table2}
\end{table}

Table~\ref{table2} shows the averaged result of 100 Monte Carlo simulations regarding tracking performance. $\bar{e}_1$, $\bar{e}_2$ and $\bar{e}_3$ in this table denote the average absolute errors for robust control with fixed set, vanilla set learning, and active learning, respectively. We calculate the error for single simulation by $e=\frac{1}{T}\sum_{k=1}^T |y(k)-y_r(k)|$ from $T=2$ to $T=10$, which evaluates the tracking performance during the start-up period. From $T=15$ to $T=45$, we calculate the error as $e=\frac{1}{T-10}\sum_{k=11}^T |y(k)-y_r(k)|$, which quantifies the tracking performance after the learning process converged. The average error for $M$ times Monte Carlo simulations is calculated as $\bar{e}=\frac{1}{M}\sum_{i=1}^M e(i)$. We use ratio to represent the performance improvement in this table, where ratio1 is defined as $(\bar{e}_1-\bar{e}_2)/\bar{e}_2$, ratio2 is $(\bar{e}_1-\bar{e}_3)/\bar{e}_3$, and ratio3 is $(\bar{e}_2-\bar{e}_3)/\bar{e}_3$. Notice that the tracking error under robust control with vanilla set learning is less than robust control with fixed set when considering the whole simulation horizon. We can observe that the tracking error under robust control with active learning is larger than the vanilla set learning and fixed set during the start-up period, while active learning over-performs the other two after $10$-th iteration. As implied by our performance improvement criteria ratio3, compared with the vanilla set learning, there is some enhancement by active learning in the tracking control, though not significant. That is, the uncertain parameter can be learned well by both vanilla set learning and active learning in this case.

\subsection{Simulation on a SISO system with three-dimensional uncertainty}\label{subsection2}

\begin{figure}[htbp]
	\centering
	\includegraphics[width=0.75\textwidth]{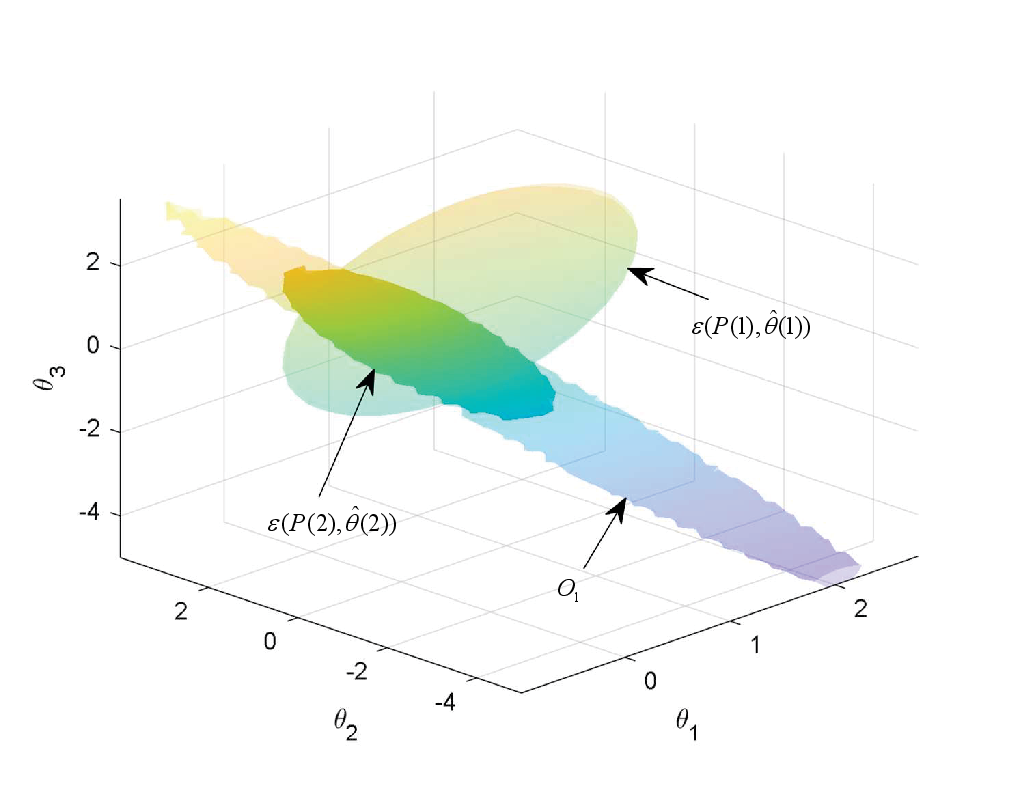}\\
	\caption{Result of the learned three dimension ellipsoidal set at the $1$-th iteration $\mathcal{E}(\bm{P}(2),\hat{\bm{\theta}}(2)) \supseteq \mathcal{E}(\bm{P}(1),\hat{\bm{\theta}}(1)) \cap O_1$. }\label{ellipsoidinter3dim}
\end{figure}

This simulation considers the same tracking control problem in subsection \ref{subsection1} while using a different state matrix 
\begin{equation}
	\begin{aligned}
		\bm{A} &= \bm{A}_0 + \alpha_1\bm{A}_1 \\
			   &= \left[ \begin{array}{ccc} 0 & 1 & 0 \\ 0 & 0 & 1 \\ -0.3 & 0.4 & 0.2 \end{array} \right] + \alpha_1 \left[ \begin{array}{ccc} 0 & 0 & 0 \\ 0 & 0 & 0.5 \\ 0 & 0 & 0 \end{array} \right],       \\
	\end{aligned} 
\end{equation}
where the parameter $\alpha_1$ is unknown. This allows us to add further uncertainty compared with the last simulation. Also note the target system output trajectory is set as a step-function wave, derived by the following function 
\begin{equation}
	\begin{aligned}
		y_r(k) = \left\lbrace  \begin{array}{rl}
			4, & k=1,2,\cdots,24 \\ -5, & k=25,26,\cdots,74 \\ -1.5, & k=75,76,\cdots,100.\\ 
		\end{array} \right. 
	\end{aligned} 
\end{equation}

In one-time simulation, the true value of parameter $\bm{\theta}$ is set as $\bm{\theta}^{\ast} = [0.15, 0.2, 0.1]^T$. The initial uncertainty ellipsoidal set is $\mathcal{E}(\bm{P}(1),\bm{\theta}(1))$, where the matrix and center are respectively set as
\begin{equation}
	\begin{aligned}
		\bm{P}(1) = \left[ \begin{array}{ccc} 3 & 1 & 1 \\ 1 & 4 & 1 \\ 1 & 1 & 2 \end{array} \right], \quad \bm{\theta}(1) = \left[ \begin{array}{c} 0.9 \\ 0.8 \\ 0.7  \end{array} \right].
	\end{aligned} 
\end{equation}
The active learning coefficient for our proposed method is set as $M_a = 0.6$.

Fig.~\ref{ellipsoidinter3dim} shows the learning result for the 3-dimensional ellipsoidal set at the $1$-th iteration by the vanilla set learning approach. The ellipsoid is learned as the intersection between the observation $O_1=\mathcal{E}(\bm{\phi}^{-1}(1)\bm{Q}(1)\bm{\phi}^{-T}(1) \bm{\phi}^{-1}(1)\bm{H}(2))$ and the prior ellipsoidal set $\mathcal{E}(\bm{P}(1),\hat{\bm{\theta}}(1))$, which we present as $\mathcal{E}(\bm{P}(2),\hat{\bm{\theta}}(2)) \supseteq \mathcal{E}(\bm{P}(1),\hat{\bm{\theta}}(1)) \cap O_1$.

\begin{figure}[htbp]
	\centering
	\includegraphics[width=0.75\textwidth]{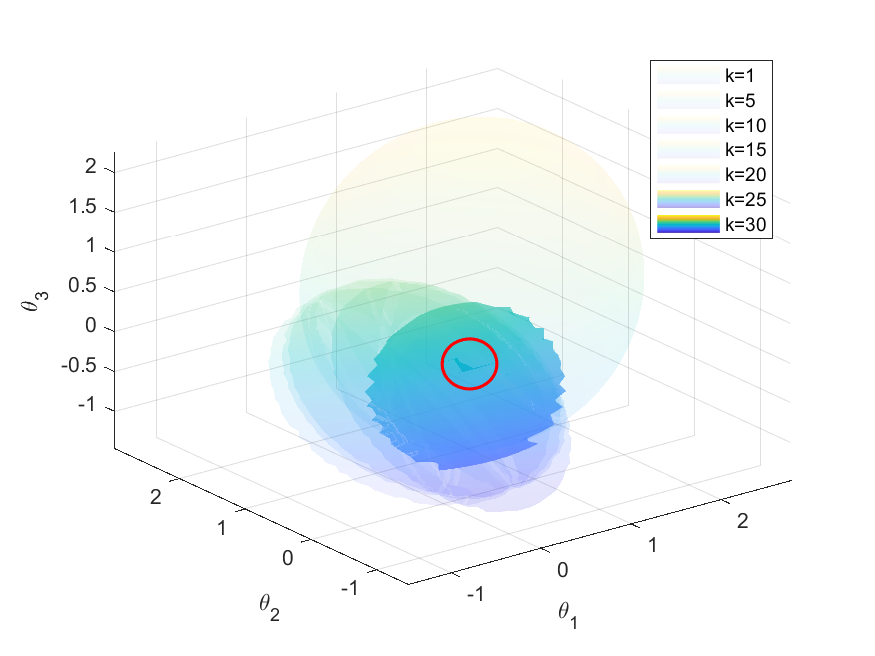}\\
	\caption{The evolution of $\mathcal{E}(\bm{P}(k),\bm{\theta}(k))$ under vanilla set learning. We circle the area of the learned ellipsoid at the last iteration in red.}\label{ellipsoidsetl3dim}
\end{figure}

\begin{figure}[htbp]
	\centering
	\includegraphics[width=0.7\textwidth]{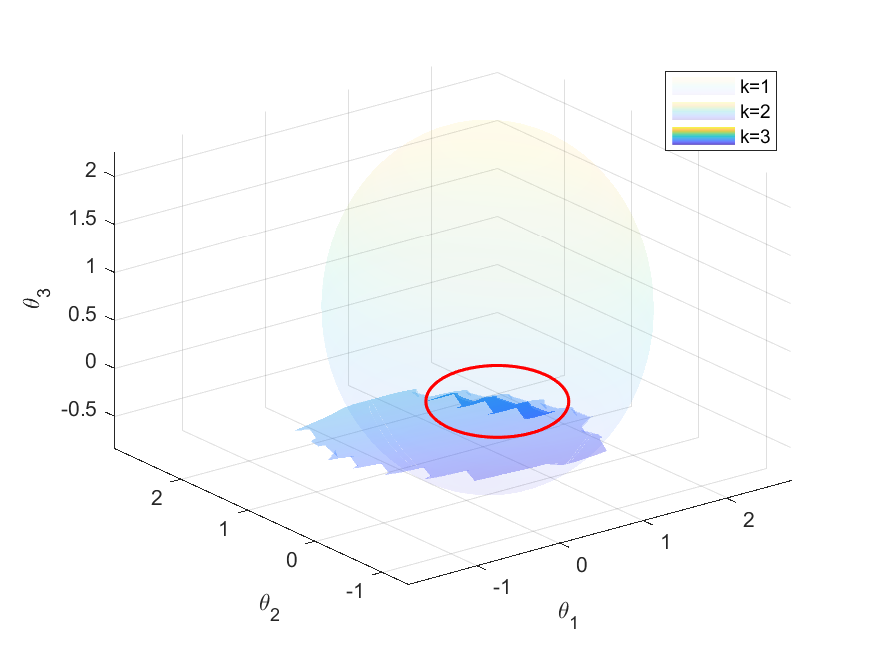}\\
	\caption{The evolution of $\mathcal{E}(\bm{P}(k),\bm{\theta}(k))$ under active learning. We circle the area of the learned ellipsoid at the last iteration in red.}\label{ellipsoidactive3dim}
\end{figure}

Fig.~\ref{ellipsoidsetl3dim} displays how the learned ellipsoidal set evolves from $1$-th to $30$-th iteration for vanilla set learning, and Fig.~\ref{ellipsoidactive3dim} shows the learned set by the proposed active learning from $1$-th to $3$-th iteration. We observe from the comparison between those two figures that the ellipsoid by active learning shrinks faster than vanilla set learning.

\begin{figure}[htbp]
	\centering
	\includegraphics[width=0.7\textwidth]{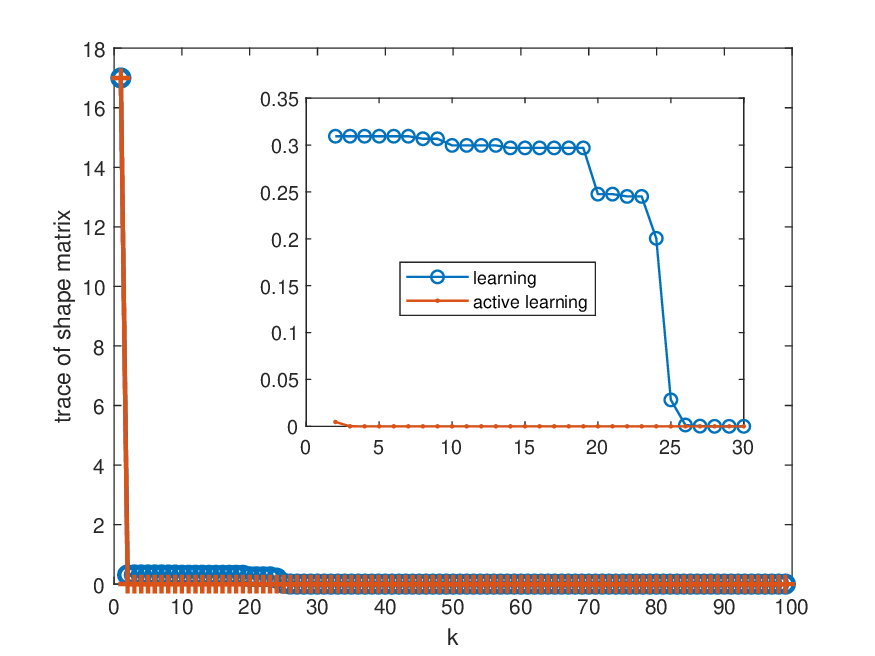}\\
	\caption{The trace of shape matrix for uncertainty ellipsoidal set with set learning and active learning.}\label{converge3dim}
\end{figure}

To quantify and visualize the uncertainty over the control horizon, Fig.~\ref{converge3dim} plots the trace of the shape matrix of the ellipsoidal set. We observe that the trace under both vanilla set learning and active learning is reduced and converges to $0$ over time. In comparison, the volume with active learning approximates $0$ after $3$ iterations, while it is $26$ iterations for vanilla set learning, demonstrating an acceleration of set learning by the designed active learning.

\begin{figure}[htbp]
	\centering
	\includegraphics[width=0.7\textwidth]{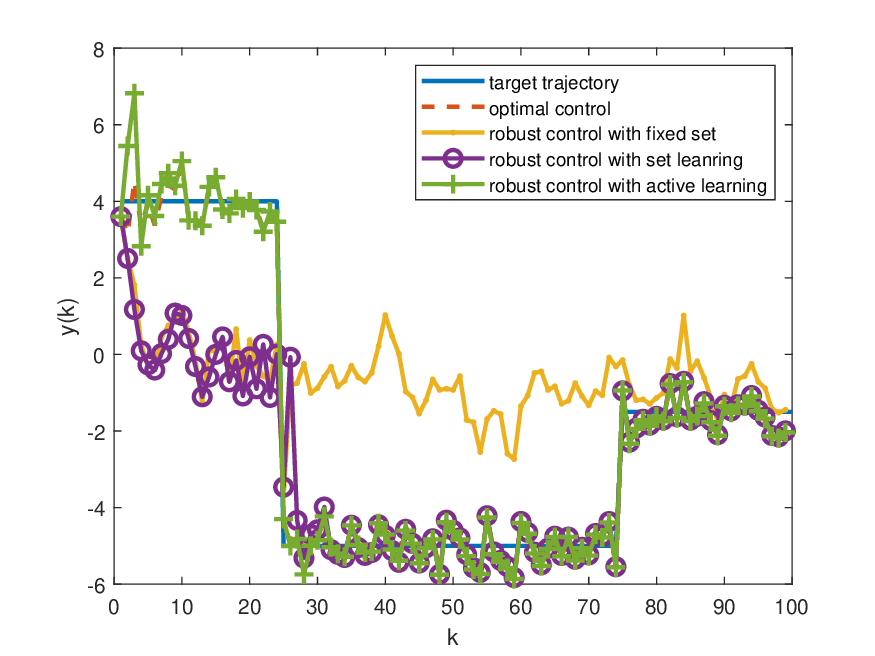}\\
	\caption{The system output target trajectory tracking signal for optimal control, robust control with fixed set, robust control with set learning and robust control with active learning.}\label{control3dim}
\end{figure}

Fig.~\ref{control3dim} shows the system output under different control approaches for target trajectory tracking. We adopt the same four control approaches as in section~\ref{subsection1}. As we can observe from this figure, compared with robust control with fixed uncertainty set and set learning, the proposed robust control with active learning is more approximate to optimal control than robust control with either fixed uncertainty set or vanilla set learning. Our approach shows a more accurate trajectory tracking except for a short spike in the start-up period. I.e., active learning brings excitation leading to a spike during the first $3$ iterations. The excitation is intended to stimulate the system to acquire more information to feed the uncertainty learning, which is supposed to bring better learning performance and enhanced tracking performance in the consequent iterations. The result actually corroborates our argument that, the robust control with active learning approximates the ideal benchmark within the first $10$ iterations after the spike in the beginning $3$ iteration, while it takes $30$ iterations for the robust control with vanilla set learning to approximate the ideal benchmark.

Table~\ref{table3} presents the averaged results of 100 Monte Carlo simulations for trajectory tracking control. The definitions of $e$ and ratio are the same as in section~\ref{subsection1}. Notice that the tracking performance under both vanilla set learning and active learning has about $600\% -800\%$ improvement after the $30$-th iteration compared with the robust control with fixed set. Significantly, from $5$-th to $30$-th iteration, the tracking error under robust control with active learning is $100\%-300\%$ and $100\%- 150\%$ reduced than fixed set and vanilla set learning approach, respectively. After the $30$-th iteration, the tracking performance under active learning is almost the same as that of the vanilla set learning. Overall, the designed active learning brings higher learning quality in fewer iterations, leading to more accurate system output tracking than the vanilla set learning and fixed uncertainty set approach.

\begin{table}[H]
	\caption{Comparison of trajectory tracking between robust control with fixed set, vanilla set learning, and active learning set.}
	\centering
	\resizebox{10cm}{!}
	{\begin{tabular}{ccccccccc}
			\toprule
			T &  $e_1$ & $e_2$ &  $e_3$ & ratio1 & ratio2 & ratio3\\ 
			\midrule
			2  & 0.5191 & 0.5191 & 0.9939 & 0.00\%     & -47.76\%  & -47.76\%\\ 
			5  & 2.1334 & 1.8293 & 0.8896 & 16.62\%  & 139.81\%  & 105.62\%\\
			10 & 2.9868 & 2.2929 & 0.8672 & 30.26\%  & 244.42\%  & 164.40\%\\
			20 & 3.3328 & 2.2978 & 0.8834 & 45.03\%  & 277.26\%  & 160.11\%\\
			30 & 3.4878 & 1.9933 & 0.8438 & 74.97\%  & 313.34\%  & 136.23\%\\
			\midrule
			40 & 4.0515 & 0.4445 & 0.4524 & 811.52\% & 795.62\% & -1.74\%\\
			55 & 4.0802 & 0.4223 & 0.4411 & 866.12\% & 825.05\% & -4.25\% \\
			70 & 4.0798 & 0.4191 & 0.4423 & 873.40\% & 822.36\% & -5.24\% \\
			85 & 3.4275 & 0.4126 & 0.4355 & 730.78\% & 686.99\% & -5.27\% \\
			100 & 2.8741 & 0.4059 & 0.4262 & 608.15\% & 574.36\% & -4.77\% \\
			\bottomrule
	\end{tabular}}
	\label{table3}
\end{table}

\subsection{Simulation on a MIMO system with four-dimensional uncertainty}\label{subsection3}

This section considers the tracking control of a transport aircraft model. We take the aircraft longitudinal dynamics and the problem setup for the aircraft control as in~\cite{eugene2013robust}, and the continuous-time model for the aircraft is:

\begin{equation}\label{longit}
	\begin{aligned}
		\left( \begin{array}{c} \dot{V} \\ \dot{\alpha} \\ \dot{q} \\ \dot{\theta}  \end{array} \right) &= \left( \begin{array}{cccc} -0.038 & 18.984 & 0 & -32.174 \\ -0.001 & -0.632 & 1 & 0 \\ 0 & -0.759 & -0.518 & 0 \\ 0 & 0 & 1 & 0 \end{array}  \right) \left( \begin{array}{c} V \\ \alpha \\ q \\ \theta  \end{array} \right)\\
		&+\left( \begin{array}{cc} 10.1 & 0 \\ 0 & -0.0086 \\ 0.025 & -0.011 \\ 0 & 0 \end{array}  \right) \left( \begin{array}{c} \delta_{th} \\ \delta_e  \end{array} \right)+\bm{\omega},\\
	\end{aligned}
\end{equation}

with the aircraft trimmed at $V_0 = 250 $ft/s and flying at a low altitude. In this model, the system state $V$ denotes the airspeed, $\alpha$ is the angle of attack, $q$ is the pitch rate, $\theta$ is the pitch angle, and the control signal $\delta_{th}$ is the airspeed control, $\delta_e $ is the pitch control. We use the zero order hold method to discretize the model with a 0.1s sampling interval, and obtain the discretized model as (\ref{sys1}), where the dynamics are presented as
\begin{equation}\label{longitDyna}
	\begin{aligned}
		&\bm{A} = \left[ \begin{array}{cccc} 0.9962 & 1.8984 & 0 & -3.2174 \\ -0.0001 & 0.9368 & 0.1 & 0 \\ 0 & -0.0759 &  0.9482 & 0 \\ 0 & 0 & 0.1 & 1 \end{array} \right]  ,\\
		&\bm{B} = \left[ \begin{array}{cc} 1.01 & 0 \\ 0 & -0.0009 \\ 0.0025 & -0.0011 \\ 0 & 0 \end{array}  \right] .\\
	\end{aligned}
\end{equation}

Assume that the dynamics $\bm{A}$ and $\bm{B}$ are the nominal matrices $\bm{A}_0$ and $\bm{B}_0$, respectively. We set that the dynamics have an uncertainty of $10\%$ in the nominal values $\bm{A}(1,1)$, $\bm{A}(1,2)$, $\bm{B}(1,1)$, and $\bm{B}(2,2)$. Therefore, we obtain the matrices $\bm{A}_1$, $\bm{A}_2$, $\bm{B}_1$ and $\bm{B}_2$ as
\begin{equation}\label{longitDyna12}
	\begin{aligned}
		&\bm{A}_1 = \left[ \begin{array}{cccc} 0.09962 & 0 & 0 & 0 \\ 0 & 0 & 0 & 0 \\ 0 & 0 & 0 & 0 \\ 0 & 0 & 0 & 0 \end{array} \right],  
		\bm{B}_1 = \left[ \begin{array}{cc} 0.101 & 0 \\ 0 & 0 \\ 0 & 0 \\ 0 & 0 \end{array}  \right], \\
		&\bm{A}_2 = \left[ \begin{array}{cccc} 0 & 0.18984 & 0 & 0 \\ 0 & 0 & 0 & 0 \\ 0 & 0 & 0 & 0 \\ 0 & 0 & 0 & 0 \end{array} \right], 
		\bm{B}_2 = \left[ \begin{array}{cc} 0 & 0 \\ 0 & 0.00086 \\ 0 & 0 \\ 0 & 0 \end{array}  \right]. \\		
	\end{aligned}
\end{equation}

With the settings above, this simulation aims to design the autopilot to track a sinusoidal angle of attack and a constant airspeed. We achieve this via applying and comparing the robust control with fixed uncertainty set, robust control with vanilla set learning, and our designed robust control with active learning. For this task, we have the observation matrix $\bm{C}$ as 
\begin{equation}
	\begin{aligned}
		\bm{C} = \left[ \begin{array}{cccc} 1 & 0 & 0 & 0\\ 0 & 1 & 0 & 0   \end{array} \right].
	\end{aligned} 
\end{equation}
We set the target airspeed and angle of attack by the function 
\begin{equation}
	\begin{aligned}
		\bm{y}_r(k) =  \left[  \begin{array}{c}	250 \\ 6\sin(k\pi/25) \end{array} \right] ,  k=1,2,\cdots,100.
	\end{aligned} 
\end{equation}
The disturbances $\bm{\omega}(k)$ are confined in an ellipsoid and we set the shape matrix as
\begin{equation}
	\begin{aligned}
		\bm{R} = \left[ \begin{array}{cccc} 0.1 & 0 & 0 & 0\\ 0 & 0.05 & 0 & 0\\  0 & 0 & 0.05 & 0\\  0 & 0 & 0 & 0.05 \end{array} \right].\\
	\end{aligned} 
\end{equation}

In one-time simulation, we set the true value of parameter $\bm{\theta}$ as $\bm{\theta}^{\ast} = [0.5, -0.8, 0.2, -0.6]^T$. The initial state is set as $\bm{x}(:,1) = [250, 0.01, 0.01, 0.01]^T$. The initial uncertainty ellipsoidal set is $\mathcal{E}(\bm{P}(1),\bm{\theta}(1))$, where the matrix and center are respectively set as
\begin{equation}
	\begin{aligned}
		&\bm{P}(1) = 2\bm{I}_{4\times4}, \\
		& \bm{\theta}(1) = \left[ \begin{array}{cccc} 0.1 & -0.1 & 0.1 & -0.1  \end{array} \right]^T.
	\end{aligned} 
\end{equation}

The active learning coefficient matrix for our proposed method is set as

\begin{equation}
	\begin{aligned}
		\bm{M}_a = \left[ \begin{array}{cc} 0.2 & 0 \\ 0 & 10000\end{array} \right].
	\end{aligned} 
\end{equation}

\begin{figure}[H]
	\centering
	\includegraphics[width=0.68\textwidth]{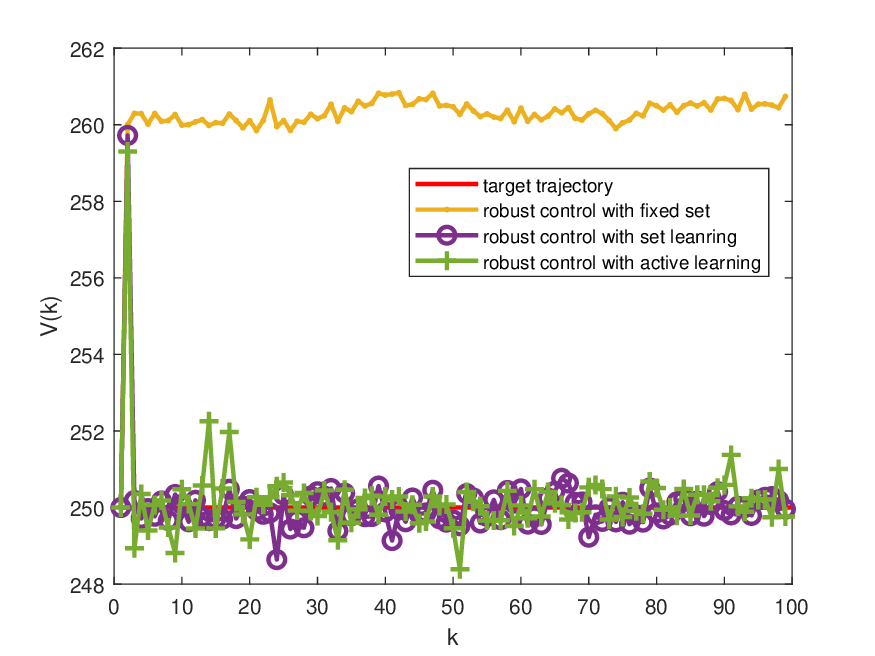}\\
	\caption{The airspeed tracking under robust control with fixed set, robust control with vanilla set learning, and robust control with active learning.}\label{TrackV}
\end{figure}

\begin{figure}[H]
	\centering
	\includegraphics[width=0.68\textwidth]{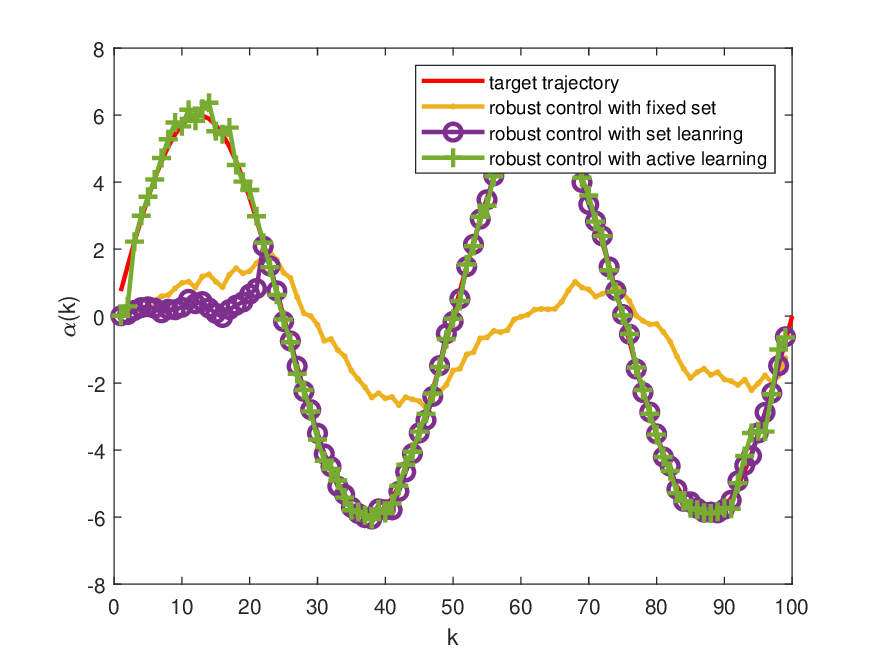}\\
	\caption{Tracking of angle of attack tracking under robust control with fixed set, vanilla set learning, and the proposed active learning.}\label{Trackalpha}
\end{figure}

Fig.~\ref{TrackV} and Fig.~\ref{Trackalpha} present the tracking performance for the airspeed and angle of attack, respectively. From Fig.~\ref{TrackV}, we can observe that robust control with either vanilla set learning or active learning can push the aircraft speed to track the target trajectory well, except for the spike from both active learning and the vanilla set learning at the beginning of the control. However, the airspeed under robust control with fixed set deviates a lot from the target trajectory throughout the control horizon. While the designed active learning and the vanilla set learning approach provide comparable tracking performance for airspeed tracking control, Fig.~\ref{Trackalpha} demonstrates the advantage of active learning over vanilla set learning. I.e., the angle of attack converges to the target trajectory within $5$ iterations under robust control with active learning, while it takes more than $20$ iterations under vanilla set learning.  

\begin{figure}[htbp]
	\centering
	\includegraphics[width=0.68\textwidth]{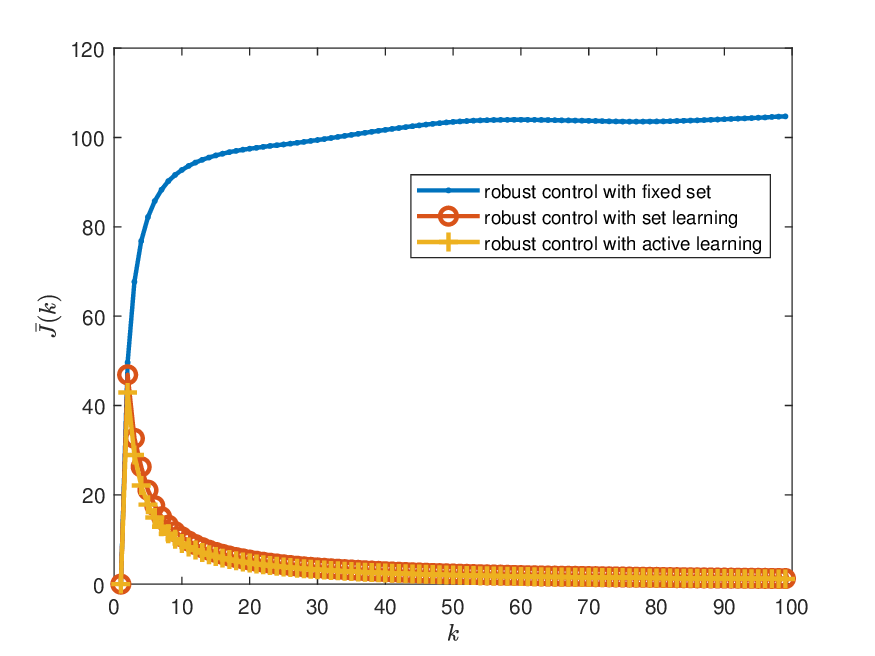}\\
	\caption{The accumulated cost for the tracking control of airspeed under robust control with fixed set, vanilla set learning, and active learning.}\label{JV}
\end{figure}

\begin{figure}[htbp]
	\centering
	\includegraphics[width=0.68\textwidth]{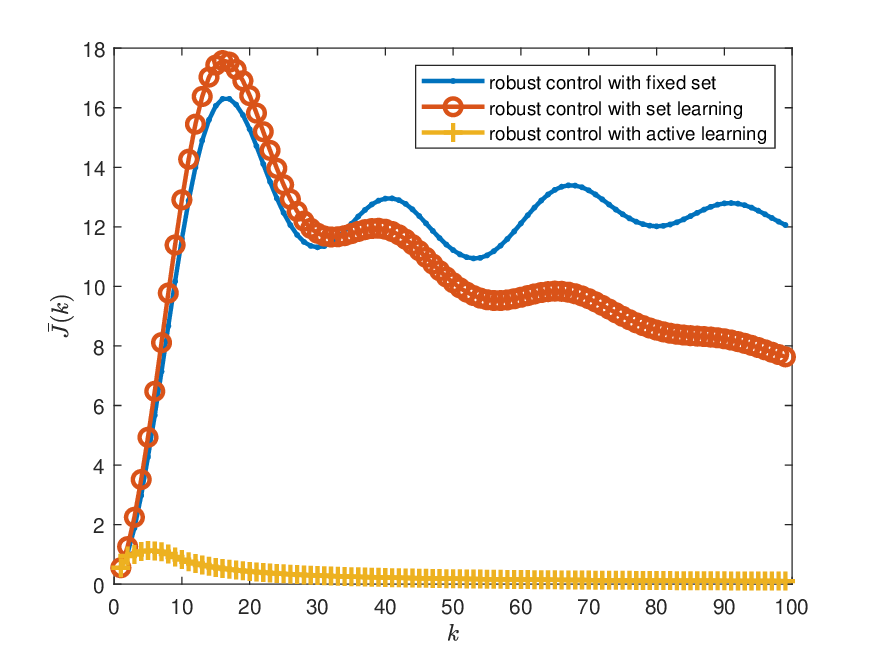}\\
	\caption{The accumulated cost for the tracking control of angle of attack under robust control with fixed set, vanilla set learning, and active learning.}\label{Jalpha}
\end{figure}

To quantify and visualize the tracking control performance, Fig.~\ref{JV} and Fig.~\ref{Jalpha} present the average accumulated cost for the tracking control under the three considered control strategies. We calculated the accumulated cost by $J(k)= 1/k \sum_{i=1}^k e^2(k)$. We define the average accumulated cost as $\bar{J}(k)=1/N_M \sum_{j=1}^{N_M} J_j(k)$, where $N_M$ is the number of Monte Carlo simulation. Fig.~\ref{JV} manifests comparable tracking performance between the control with vanilla set learning and active learning, both of which perform much better than the control with fixed uncertainty set. In Fig.~\ref{Jalpha}, the proposed approach presents a significant improvement in tracking control compared with both the control with vanilla set learning and the control with fixed uncertainty set. Such a result implies the advantage of our approach with more efficient uncertainty reduction that benefits the trajectory tracking control goal.

\section{Conclusion}\label{section7}

We proposed a novel adaptive robust control for linear uncertain systems, and endowed the control with full-fledged learning capacity toward uncertainties. We enabled the uncertainty learning via integrating ellipsoidal set-membership estimation into the design of robust control. We reformulated the robust control law derivation and provided a solution that is approximate to optimal, along with computational feasibility. We enhanced the uncertainty learning through ellipsoid set volume maximization, and connected this uncertainty learning with robust control law derivation. We solved the conflict between control and learning in the derived robust control with active learning, leading to a control strategy that can learn the uncertainties from enriched sources of information. We conducted simulations using diversified settings to verify our approach and make comparisons, and the results demonstrated the advantage of the proposed approach with more accurate and accelerated uncertainty learning and higher tracking control performance.

An interesting direction for relevant future research is to extend the method to the control of nonlinear uncertain systems, where the main challenge can lie in the modification of ellipsoidal set-membership estimation to fit nonlinear systems. Furthermore, we plan to apply the proposed adaptive robust control to practical system control to testify and improve this method further.

\bibliographystyle{IEEEtran}
\bibliography{bibfile}

\end{document}